\documentclass[prd,aps, nofootinbib,singlecolumn]{revtex4-2}

\usepackage{amsmath,amssymb,amsfonts,color,graphicx,graphics,latexsym,placeins,epsfig,multirow}
\usepackage[toc]{appendix}
\usepackage{array}
\newcolumntype{P}[1]{>{\centering\arraybackslash}p{#1}}
\usepackage{footmisc}
\usepackage{mathrsfs}
\usepackage[compatibility=false]{caption}
\usepackage{subcaption}
\usepackage{comment}
\usepackage{bbold}
\usepackage{enumitem}
\setdescription{leftmargin=12.5pt}
\usepackage{bbold}
\usepackage{hyperref}
\usepackage{varioref}
\usepackage{color}
\usepackage{float}
\begin{document}

\pdfoutput=1

\newcommand{\sn}{{\rm sn}}
\newcommand{\cn}{{\rm cn}}
\newcommand{\dn}{{\rm dn}}
\newcommand{\sech}{{\rm sech}}
\newtheorem{theorem}{Theorem}[section]
\newtheorem{lemma}[theorem]{Lemma}

\newtheorem{definition}[theorem]{Definition}
\newtheorem{example}[theorem]{Example}
\newtheorem{xca}[theorem]{Exercise}
\newcommand{\lso}[2]{#1\left(\textcolor{#2}{^{\line(1,0){20}}}\right)}
\newcommand{\lda}[2]{#1\left(\textcolor{#2}{^{_{\bf{------}}}}\right)}
\newcommand{\loo}[2]{#1\left(\textcolor{#2}{^{_{\bf{ooo}}}}\right)}
\newtheorem{remark}[theorem]{Remark}

\def\a{\alpha}
\def\b{\beta}
\def\c{\gamma} 
\def\d{\delta}
\def\e{\epsilon}           % Also, \varepsilon
\def\f{\phi}               %      \varphi
\def\vf{\varphi}  \def\tvf{\tilde{\varphi}} 
\def\g{\gamma}
\def\h{\eta}   
\def\i{\iota}
\def\j{\psi}
\def\k{\kappa}                    % Also, \varkappa (see below)
\def\l{\lambda}
\def\m{\mu}
\def\n{\nu}
\def\o{\omega}  \def\w{\omega}
\def\p{\pi}                % Also, \varpi
\def\q{\theta}  \def\th{\theta}                  %     \vartheta
\def\r{\rho}                                     %     \varrho
\def\s{\sigma}                                   %     \varsigma
\def\t{\tau}
\def\u{\upsilon}
\def\x{\xi}
\def\z{\zeta}
\def\D{\Delta}
\def\F{\Phi}
\def\G{\Gamma}
\def\J{\Psi}
\def\L{\Lambda}
\def\O{\Omega}  \def\W{\Omega}
\def\P{\Pi}
\def\Q{\Theta}
\def\S{\Sigma}
\def\U{\Upsilon}
\def\X{\Xi}
\def\del{\partial}              
% overwritten by \nabla

% Calligraphic letters

\def\ca{{\cal A}}
\def\cb{{\cal B}}
\def\cc{{\cal C}}
\def\cd{{\cal D}}
\def\ce{{\cal E}}
\def\cf{{\cal F}}
\def\cg{{\cal G}}
\def\ch{{\cal H}}
\def\ci{{\cal I}}
\def\cj{{\cal J}}
\def\ck{{\cal K}}
\def\cl{{\cal L}}
\def\cm{{\cal M}}
\def\cn{{\cal N}}
\def\co{{\cal O}}
\def\cp{{\cal P}}
\def\cq{{\cal Q}}
\def\car{{\cal R}}
\def\cs{{\cal S}}
\def\ct{{\cal T}}
\def\cu{{\cal U}}
\def\cv{{\cal V}}
\def\cw{{\cal W}}
\def\cx{{\cal X}}
\def\cy{{\cal Y}}
\def\cz{{\cal Z}}

%%%%%%%%%%%%%%%%%%%%%%%%%%%%%%%%%%%%%%%%%%%%%%%%%%%%%%%%%%%%%

% Umut needs

%\def\bd{\begin{displaymath}}
%\def\ed{\end{displaymath}}
%\def\quart{\frac14}
\def\6{\partial}
%\def\N4{{\cal N}=4}
%\def\lab{\label}
%\def\bq{\bar{q}}
%%%%%%%%%%%%%%%%%%%%%%%%%%%%%%%%%%%%%%%%%%%%%%%%%%%%%%%%%%%%%%

\title{\Large \bf Gravitational wave memory for a class of static and spherically symmetric spacetimes }
\author{Soumya Bhattacharya$^{1}$ }
\author{Shramana Ghosh$^{2,3}$}
%\author{Z}
\affiliation{$^{1}$ Department of Astrophysics and High Energy Physics, S.N. Bose National Center for Basic Sciences, Kolkata 700106, India}
\email{soumya557@bose.res.in}

\affiliation{$^{2}$ School of Physical Sciences, Indian Association for the Cultivation of Science, Kolkata 700032, India}
\email{shramana.ghosh.kolkata@gmail.com}
\affiliation{$^{3}$  Department of Physics, Montana State University, Bozeman, Montana-59715}
\email{shramanaghosh@montana.edu}
%\fntext[myfootnote1]{Email: soumya557@bose.res.in (Soumya Bhattacharya)}

\begin{abstract}
This article aims at comparing gravitational wave memory effect in a Schwarzschild spacetime with that of other compact objects with static and spherically symmetric spacetime, with the purpose of proposing a procedure for differentiating between various compact object geometries. We do this by considering the relative evolution of two nearby test geodesics with  in different backgrounds in the presence and absence of a gravitational wave pulse and comparing them. Memory effect due to a gravitational wave would ensure that there is a permanent effect on each spacetime and the corresponding geodesic evolution, being metric dependent, would display distinct results in each case. For a complete picture, we have considered both displacement and velocity memory effect in each geometry.
\end{abstract}

\maketitle
\section{Introduction} \label{sec1}
Einstein's theory of General Relativity has been the most successful theory of gravity till date. However, it fails to explain with certainty regions of extreme gravity, for example: in the vicinity of singularities and the starting point of the universe itself. With the rise of observational gravitational wave (GW) data from detectors like LIGO \cite{LIGO1, LIGO2}  as well as the observations of the shadow of
supermassive compact central objects e.g., the M87* and
the SgrA* by the Event Horizon Telescope (EHT) \cite{EHT1, EHT2, EHT3, EHT4, EHT5, EHT6, EHT7}, we now have the means necessary to propose and possibly check alternative theories of gravity by looking for higher dimensions or changes in the known structure or behaviour of black hole spacetime. Such predictions of black hole like compact objects have lead to new studies in Exotic Compact Objects (ECOs) \cite{Yunes, LIGO-VIRGO, Ohme, Genzel, Yunes1, psaltis, SC1, SC2, SC3, SC4} arising from theories of quantum fluctuations and/or dark matter. These objects behave as black holes for solar system tests of gravity but they may display distinct features when probed using strong field tests of gravity, such as gravitational waves.
Although black holes are a widely studied exact solution of Einstein's field equations and we have a huge inventory of data that points to their existence, we still cannot say for sure because of the lack of experimental/observational evidence of the event horizon which is a defining feature of any black hole. Hence it is interesting to study to what extent the current or upcoming experiments or detectors can observationally establish the existence of black holes given that there are so many possible black hole mimickers. 
The first image released of the supermassive  black hole at the centre of M87*, an elliptical galaxy did not just boost research in the realm of black hole physics but also gave rise to a very fundamental question,i.e. in the absence of any proof of an event horizon is it really a black hole? Hence, although black hole research is a prominent field of study today, given the ever increasing volume of gravitational wave data, this particular question arises because all that we can verify is the existence of a photon sphere. However, the proof of photon sphere alone is not enough for the existence of a black hole. In the darkness beyond a photon sphere, many postulate that other exotic compact objects may exist, although their stability studies are not at par with those of a black hole. \\
There are two ways in which this question can be resolved - either we attempt to prove the existence of the horizon or we find some procedure by which we can differentiate between various compact object geometries. In this work, we try to pursue the latter path and present a comparative study between various ECO geometries. We take gravitational wave {\it memory effect} as our measurable phenomenon and consider static and spherically symmetric solutions of various models of wormholes and other theories of gravity for comparison with a simple Schwarzschild black hole. As these regions of extreme gravity are perfect laboratories for studying higher dimensional theories of gravity hence we consider those models as well. 
However, we must warn that current gravitational wave detectors like LIGO do not have the sensitivity required to detect such a minute difference caused due to memory effect. To demonstrate this we try to find the order of strain sensitivity required to detect memory effect with the current strain sensitivity of LIGO. But we do expect next generation of detectors to be able to detect gravitational wave memory effect which will enable us to realise this study in the experimental front. The improvement in the detection prospects for future ground based GW detectors and also the launch of the space based-detector LISA in the near future, may provide us an opportunity to observe GW {\it memory effect} \cite{Nichols1}. This effect refers to the lasting change in the relative distance between
test particles when a GW passes through the ambient
spacetime \cite{braginsky, favata}. The memory effect encompasses both the strong-field, as well as non-linear aspects of general relativity, which is yet to be observed. \\
\noindent Keeping in mind the above importance  we study memory effect for various spherically symmetric spacetimes of GR and theories beyond GR. We analyse the memory effect here by studying the geodesic evolution of two test particles both in the absence and in the presence of a gravitational wave and making a comparative study. We believe this study will shed some important light on our understanding of memory effect as well as some broad aspects like the strong gravity regime of GR and theories beyond GR. \\
\noindent This paper is organized as follows:
First we introduce some basic features of static, spherically symmetric geometry in section \ref{sec2}. We then study the gravitational wave memory for various static and spherically symmetric black hole geometries in section \ref{sec3}. In section \ref{sec4}, gravitational wave memory have been studied for various static, spherically symmetric wormhole solutions which are possible candidates of black hole mimickers. Then we make a comparative study of gravitational wave memory in section \ref{sec5}. Finally, concluding remarks have been given in section \ref{sec6}. 
%memory effect and how it can be used to detect the passing of a gravitational wave by demonstrating it in a Schwarzschild background \ref{sec1}. 
%Next we perform the same procedure for a variety of other spacetime geometries that can be classified as black hole mimickers, categorised into two sections: Static and Spherically symmetric wormhole geometries\ref{sec4} and  With all the obtained data, we finally compare\ref{sec4} memory effect in Schwarzschild geometry with memory effect in all other spacetime geometries and categorise them as spacetimes which might show  memory effect greater than or less than a Schwarzschild black hole. %Thus, using this evidence we conclude that memory effect, if ever detected, might be an important tool in differentiating between different geometries in the future. We also briefly mention if LIGO can detect memory effect and compare the strain sensitivity required.
\section{Geodesic evolution in static and spherically symmetric spacetime in the presence and absence of Gravitational Waves} \label{sec2}
\noindent In this section, we briefly discuss the geodesic equations for any static and spherically symmetric spacetime in presence and absence of gravitational waves. The general equations we get here will be used in later sections to describe memory effect for various static, spherically symmetric backgrounds. So first let us write down the line element that describes any static and spherically symmetric spacetime
\begin{equation}
 ds^2=-f(r)dt^2 + \frac{dr^2}{g(r)} + r^2d\Omega_2 ^2  \label{sphm} 
\end{equation} 
\noindent If $f(r) = 0$ at some value of $r=r_H$ then the geometry described by eqn \ref{sphm} represents a black hole. However if there is no horizon, or singularity present in the solution, and the solution is asymptotically flat then the solution represents a geometry featuring
two asymptotic regions connected by a bridge or a wormhole solution. We now consider the geodesic equations for this metric. We know for arbitrarily general spacetime, described by spacetime coordinates $x^a$, the geodesic equation for a free-falling object in this spacetime can be constructed as follows
\begin{equation}
  \frac{d^2 x^a}{d \tau^2} + \Gamma^{a}_{~bc} \frac{dx^b}{d\tau}\frac{dx^c}{d\tau} = 0   
\end{equation}
Where $\Gamma^{a}_{~bc}$ are the affine connections corresponding to the arbitrary spacetime geometry and $\tau$ is the affine parameter. 
So the geodesic equations corresponding to the line element \ref{sphm} look like the following (considering equatorial plane $\theta = \pi/2$):
\begin{align}
\ddot{t} +  \frac{f'(r)}{f(r)} ~\dot{r} \dot{t}&=0 \label{eg1}\\ 
   \ddot{r} -\frac{g'(r)}{2g(r)}~\dot{r}^2+\frac{f'(r)g(r)}{2}~\dot{t}^2-rg(r)~\dot{{\phi}}^2 &=0 \label{eg2}\\ 
   \ddot{\phi} + \frac{2}{r}~\dot{r} \dot{\phi}&=0 \label{eg3}
\end{align}
The convention we follow is: dot represents derivative with respect to proper time $\tau$ and dash represents derivative with respect to the associated coordinate, for example $ f'(r)= df/dr$ and $\dot r=dr/d\tau$. 
To solve these geodesic equations, we need to specify initial conditions which will be determined by the velocity normalisation condition.
\begin{equation}
    g_{ab}u^{a}u^{b}=-1 \label{vel norm}
\end{equation}
which for our case reduces to
\begin{equation}
    -f(r)\dot{t}^2 + \frac{1}{g(r)}\dot{r}^2+r^2\dot{\phi}^2=-1
\end{equation}
Now, for our gravitational wave we take the pulse profile
\begin{equation}
    H(t)=A~ \sech^2(t-t_0)  \label{wave}
\end{equation}
If we consider the cross-components of the GW to be zero for simplicity then we can write the above line element in \ref{sphm} in transverse-traceless (TT) gauge as follows
\begin{equation}
 \large ds^2=-f(r)dt^2 + \frac{dr^2}{g(r)} + \left(r^2+rH(t)\right)d\theta^2 + \left(r^2-rH(t)\right)\sin^2\theta d\phi^2  \label{sphmw}
\end{equation}
Corresponding to this, the geodesic equations would be
\begin{align}
    \ddot{t}-\frac{r  H'(t)}{2 f(r)}~\dot{\phi}^2 + \frac{f'(r)}{f(r)}~\dot{r} \dot{t}&=0 \label{egw1}\\
    \ddot{r} - \frac{ g'(r)}{2g(r)}~\dot{r}^2+ \frac{ f'(r)g(r)}{2}~\dot{t}^2 +  \left(\frac{ H(t)-2r}{2} \right)g(r)~\dot{\phi}^2 &=0 \label{egw2}\\
    \ddot{\phi} + \left(\frac{2r-H(t)}{r^2-r H(t)}\right)~\dot{r} \dot{\phi} - \left(\frac{ H'(t)}{r-H(t)}\right)~ \dot{t} \dot{\phi}&=0 \label{egw3}
\end{align}
We will now make use of these equations to demonstrate memory effect in specific spacetime geometries. For standardisation purposes, we have taken $M=1$ in all our calculations and we confine our calculations in the equatorial plane i.e. $\theta = \pi/2$ for simplicity but without losing any generality. Another important point to be noted here is that the parameter $\tau$ we will use for our work is an approximate affine parameter, i.e. it remains affine only till a certain range in which the coordinate $t$ in the presence of a gravitational wave shows linear behaviour. Since we are considering only the range of $\tau$ where the gravitational wave impacts our test particle geodesic hence this approximation is fair.

\section{Memory effect in Static and Spherically symmetric Black hole solutions}\label{sec3}
\noindent In this section we consider some static spherically symmetric black hole (BH) solutions starting from Schwarzschild solution as well as some other BH solutions from theories beyond GR and explore the GW memory effects for these black hole backgrounds. 
\subsection{Memory Effect in Schwarschild spacetime}
\noindent The Schwarzschild solution is a static, spherically symmetric solution of the vacuum Einstein equations. This solution represents the black hole solution as well as the exterior geometry of any spherically symmetric gravitational source. The uniqueness of this solution is guaranteed by Birkhoff's theorem which states that any spherically symmetric solution of the vacuum field equations must be static and asymptotically flat and hence must be described by the line element \ref{sphm}, for which $f(r)$ and $g(r)$ will have the following form
\begin{equation}
    f(r) = g(r) = \left(1-\frac{2}{r}\right) \label{schw}
\end{equation}
\noindent Let us demonstrate Memory effect in the simplest case of a Schwarzschild black hole background. We follow the idea of memory effect from the work of Braginskij \cite{brag} which talks about how memory effect can be interpreted simply as a Newtonian force acting between two particles when a gravitational wave passes through them and thus integrating the force equation using the appropriate conditions shows us that both position and velocity of a particle with respect to another particle considered to be at the origin will change. Hence here we will consider two test particles in the presence and absence of a gravitational wave. In both cases, we will take the appropriate metric and solve the geodesic equations numerically to determine the trajectory of the particles. What we will emphasize on is the difference in the co-ordinates of the two particles. When a gravitational wave passes, it causes a permanent change in the co-moving distance between two particles, this is known as displacement memory effect. Hence, by tracking the relative distance between two particles as they evolve in time we expect to see that the relative co-ordinate separation between the two particles would be different in the presence of a gravitational wave as compared to that in the absence of a gravitational wave. Let us first solve the geodesic equations in the absence of a gravitational wave.
We use the velocity normalisation condition \ref{vel norm} to determine the initial conditions.
As we have three coordinates here ($t,~r$, $\phi$), we will require six initial conditions to determine the geodesic solution. We can choose any five of those initial conditions and equation \ref{vel norm} will determine the remaining initial condition such that we get a time-like geodesic curve.
The geodesic equations (\ref{eg1} - \ref{eg3}) will take the following form
\begin{equation}
\ddot{t} +  \frac{2}{r^2-2r} ~\dot{r} \dot{t}=0
\end{equation}
\begin{equation}
   \ddot{r} -\frac{\dot r^2}{r^2-2r}+\frac{1}{r^2}\left(1-\frac{2}{r}\right)~\dot{t}^2-(r-2)~\dot{{\phi}}^2 =0
   \end{equation}
   \begin{equation}
   \ddot{\phi} + \frac{2}{r}~\dot{r} \dot{\phi}=0
\end{equation}
Now, consider a gravitational wave that looks like \ref{wave}. We could have also taken any other pulse profile, for example, a Gaussian profile.
In this study we have taken $A=10$ and $t_0=9$ for our GW.
\noindent We can write the Schwarzschild line element in presence of gravitational waves as \ref{sphmw} with $f(r)$ and $g(r)$ shown in \ref{schw}. 
\noindent The geodesic equations (\ref{egw1}-\ref{egw3}) will take the following form 
\begin{equation}
    \ddot{t}+\frac{2}{r^2-2r}~\dot{r} \dot{t}+\frac{r^2 H'(t)}{2(r-2)}~\dot{\phi}^2 =0
    \end{equation}
    \begin{equation}
    \ddot{r} - \frac{\dot{r}^2}{(r^2-2r)}+ \frac{1}{r^2}\left(1-\frac{2}{r}\right)~\dot{t}^2 +  \left(\frac{ H(t)-2 r}{2} \right)\left(1-\frac{2}{r}\right)~\dot{\phi}^2 =0
    \end{equation}
    \begin{equation}
    \ddot{\phi} + \left(\frac{2r-H(t)}{r^2-r H(t)}\right)~\dot{r} \dot{\phi} + \left(\frac{ H'(t)}{H(t)-r}\right)~ \dot{t} \dot{\phi}=0
\end{equation}
\noindent We now compare the relative differences in $t, ~r$ and velocity, where velocity is defined as $v=d(\Delta r)/d\tau$, between these two test particles in the following manner
\begin{equation}
     \Delta x_{\rm withoutGW}=\Delta x_{\rm geodesic2}-\Delta x_{\rm geodesic1}
     \end{equation}
     \begin{equation}
     \Delta x_{\rm withGW}=\Delta x_{\rm geodesic2}-\Delta x_{\rm geodesic1}
\end{equation}
Here $x$ denotes any of the $t,~r$ or $v ~(=d(\Delta r)/d\tau)$ coordinates.
We then compare these co-ordinate differences by plotting them together to show what effect a gravitational wave would have on the relative geodesic evolution of two particles in this particular background. For example,we compare $\Delta t_{\rm without~GW}$ and $\Delta t_{\rm with~GW}$ depicted in red and blue respectively to demonstrate memory effect and continue the same process with the other two co-ordinate differences.

\noindent Consider the plot for $v=dr/d\tau$ against the proper time $\tau$. Here, we do see memory effect manifested as deviation in the velocity but as we asymptotically approach larger $\tau$ values, we see this deviation disappear which might be a consequence of the fact that $\tau$ is not exactly an affine parameter for larger values. 

\begin{figure}[H]
\centering
\begin{subfigure}{.5\textwidth}
  \centering
   \includegraphics[width=0.9\linewidth]{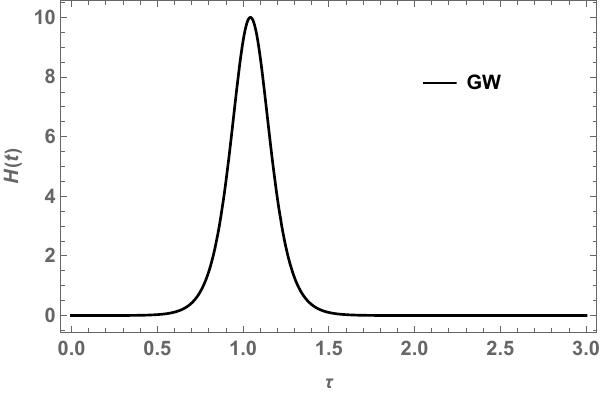}
  \caption{Gravitational wave profile}
  \label{fig:sub1}
\end{subfigure}%
\begin{subfigure}{.5\textwidth}
  \centering
   \includegraphics[width=0.9\linewidth]{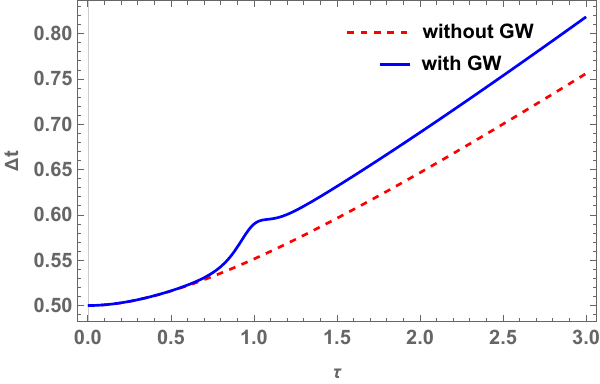}
  \caption{Deviation in $t$ co-ordinate difference}
  \label{fig:sub2}
\end{subfigure}%

\begin{subfigure}{.5\textwidth}
  \centering
  \includegraphics[width=0.9\linewidth]{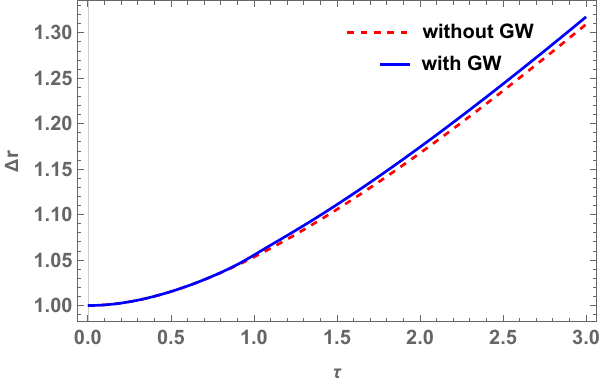}
  \caption{Deviation in $r$ co-ordinate difference}
  \label{fig:sub1}
\end{subfigure}%
\begin{subfigure}{.5\textwidth}
  \centering
  \includegraphics[width=0.9\linewidth]{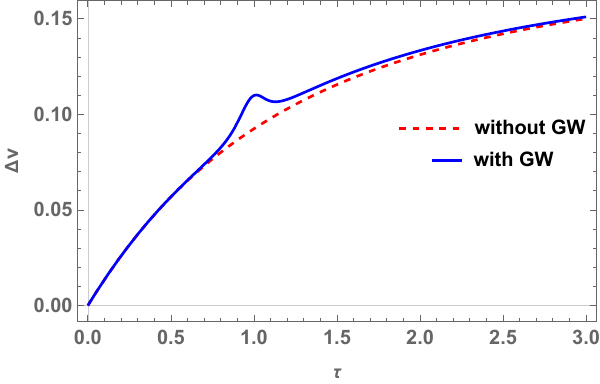}
  \caption{Deviation in velocity($v=d(\Delta r)/d\tau$) difference}
  \label{fig:sub2}
\end{subfigure}%
\caption{Memory effect in Schwarzschild spacetime}
\label{fig:test}
\end{figure}

\subsection{Braneworld Black holes}
\noindent Here we consider brane localised black holes in the Randall-Sundrum braneworld scenario which may have reflective boundary conditions arising due to quantum corrections near the horizon. As is the standard braneworld scenario, all matter exists on the four dimensional brane and only gravity can propagate through the five dimensional bulk \cite{RS1,RS2,RS3}. Thus the effective gravitational field equations can be derived using the Gauss-Codazzi formalism \cite{gauss1,gauss2,gauss3,gauss4,gauss5} and can then be solved for a static and spherically symmetric spacetime whose metric looks similar to that of a Reissner-Nordstrom black hole with its electric charge replaced by a Tidal charge $Q$. This tidal charge is essentially a manifestation of the presence of higher dimensions and thus observational indications of this tidal charge can open a window into the study of such higher dimensional theories. Although the electric charge in Reissner-Nordstrom black hole can take only positive electric charge, here the tidal charge parameter can take both positive and negative values. The Braneworld black hole metric given in \cite{braneBH} looks like \ref{sphm} with $f(r)$ and $g(r)$ will have the following form:
\begin{equation}
       f(r)= g(r) = 1-\frac{2}{r}-\frac{Q}{r^2}
\end{equation}
\noindent The overall sign of the $Q/r^2$ term in the expression for $f(r)$ determines whether it mimics a Reissner-Nordstom black hole (if the sign is positive) or if it indicates higher dimensional black hole(if the sign is negative).
This term originates from the projection of the bulk Weyl tensor which is the correction term in the effective gravitational field equations in braneworld scenario~\cite{braneBH}. The tidal charge determines how much distance the horizon penetrates into the bulk. In particular, as the tidal
charge parameter increases, the extent of the horizon in the bulk spacetime decreases, i.e., the black hole becomes more and more localized. 
    
\noindent For the given metric, in the absence of a gravitational wave, the geodesic equations are:
\begin{align}
\ddot t+\frac{2}{r}\left(\frac{r+Q}{r^2-2r-Q}\right)~\dot r \dot t&=0\\
\ddot r+\frac{(r+Q)(r^2-2r-Q)}{r^5}~\dot t^2-\frac{1}{r}\left(\frac{r+Q}{r^2-2r-Q}\right)~\dot r^2 -r\left( 1-\frac{2}{r}-\frac{Q}{r^2}\right)~\dot \phi^2&=0\\
    \ddot \phi+\frac{2 }{r}~\dot r \dot \phi&=0
\end{align}
%The velocity normalisation\ref{vel norm} condition is:
%\begin{equation}
%    -\left( 1-\frac{2}{r}-\frac{Q}{r^2}\right)\dot t^2+\frac{\dot r^2}{ 1-\frac{2}{r}-\frac{Q}{r^2}}+r^2\dot\phi^2=-1
%\end{equation}
\noindent In presence of a gravitational wave, equations (\ref{egw1}-\ref{egw3}) will take the following form 
%\begin{equation}
%    ds^2=-\left( 1-\frac{2}{r}-\frac{Q}{r^2}\right)dt^2 + \frac{dr^2}{ 1-\frac{2}{r}-\frac{Q}{r^2}} + \left(r^2+rH(t)\right)d\theta^2 + \left(r^2-rH(t)\right)\sin^2\theta d\phi^2
%\end{equation}
\begin{align}
    \ddot t+\frac{2}{r}\left(\frac{r+Q}{r^2-2r-Q}\right)~\dot r \dot t - \frac{r^3H'(t)}{2(r^2-2r-Q)} \dot \phi^2&=0\\
    \ddot r+\frac{\left(-Q+r^2-2 r\right) \dot \phi^2 (h(t)-2 r)}{2 r^2}+\frac{(Q+r) \dot r^2}{r \left(Q-r^2+2 r\right)}+\frac{(Q+r) \left(-Q+r^2-2 r\right)\dot t^2}{r^5}&=0\\
\ddot \phi +\left(\frac{2r \dot r-\dot r H(t)-rH'(t)\dot t}{r^2-rH(t)}\right )~\dot \phi &=0
\end{align}

\noindent Here, we have a charge parameter named $Q$ which can have both positive and negative values but for our study we consider only positive values.

\iffalse
\begin{figure}[H]
\begin{subfigure}{.5\textwidth}
  \centering
   \includegraphics[width=0.9\linewidth]{braneBHnegQGW.pdf}
  \caption{Gravitational wave profile}
  \label{fig:sub1}
\end{subfigure}%
\begin{subfigure}{.5\textwidth}
  \centering
   \includegraphics[width=0.9\linewidth]{braneworldBHnegQ_t.pdf}
  \caption{Deviation in $t$ co-ordinate difference}
  \label{fig:sub2}
\end{subfigure}%

\begin{subfigure}{.5\textwidth}
  \centering
  \includegraphics[width=0.9\linewidth]{braneworldBHnegQ_r.pdf}
  \caption{Deviation in $r$ co-ordinate difference}
  \label{fig:sub1}
\end{subfigure}%
\begin{subfigure}{.5\textwidth}
  \centering
  \includegraphics[width=0.9\linewidth]{braneworldBHnegQ_v.pdf}
  \caption{Deviation in velocity($v=d(\Delta r)/d\tau$) difference}
  \label{fig:sub2}
\end{subfigure}%
\caption{Memory effect in Braneworld Black Hole with $Q=-0.1$}
\label{fig:test}
\end{figure}
\fi 
\begin{figure}[H]
\begin{subfigure}{.5\textwidth}
  \centering
   \includegraphics[width=0.9\linewidth]{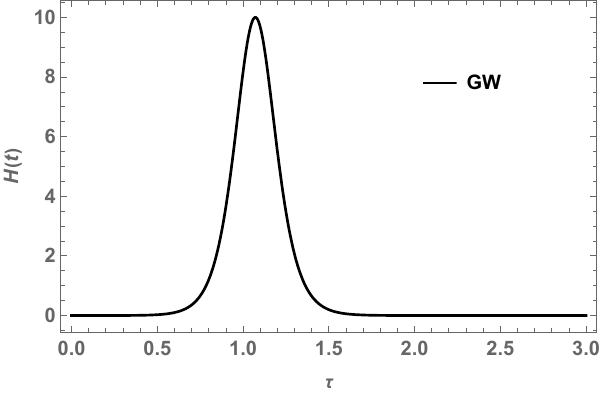}
  \caption{Gravitational wave profile}
  \label{fig:sub1}
\end{subfigure}%
\begin{subfigure}{.5\textwidth}
  \centering
   \includegraphics[width=0.9\linewidth]{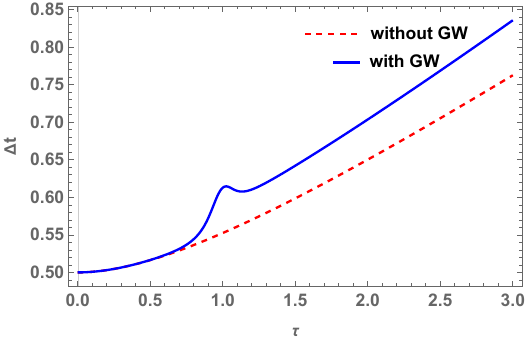}
  \caption{Deviation in $t$ co-ordinate difference}
  \label{fig:sub2}
\end{subfigure}%

\begin{subfigure}{.5\textwidth}
  \centering
  \includegraphics[width=0.9\linewidth]{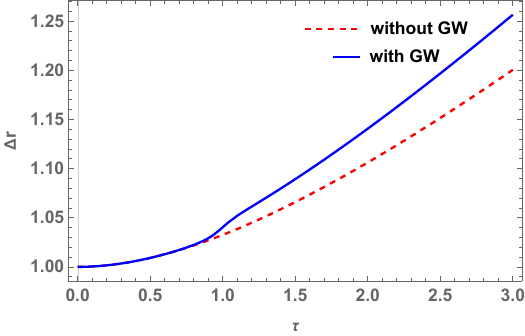}
  \caption{Deviation in $r$ co-ordinate difference}
  \label{fig:sub1}
\end{subfigure}%
\begin{subfigure}{.5\textwidth}
  \centering
  \includegraphics[width=0.9\linewidth]{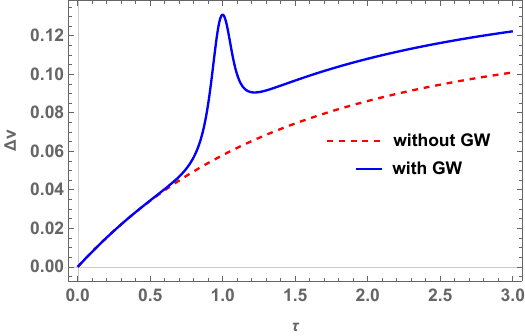}
  \caption{Deviation in velocity($v=d(\Delta r)d\tau$) difference}
  \label{fig:sub2}
\end{subfigure}%
\caption{Memory effect in Braneworld Black Hole with $Q=+0.1$}
\label{fig:test}
\end{figure}

\newpage
\subsubsection{Maeda Dadhich solution}
\noindent A static and spherically symmetric black hole solution in Einstein-Gauss-Bonnet theory of gravity in $n(>6)$ dimensional Kaluza-Klein spacetime was given by Hideki Maeda and Naresh Dadhich \cite{md}. The line element for this solution \cite{maeda} is of the form \ref{sphm}, with $f(r)$ and $g(r)$ are given by,
\begin{equation}
    f(r) = g(r) =1-\frac{2G}{r}+\frac{4G^2\tilde{q}}{r^2}
\end{equation}
We briefly mention this metric here because it is similar in form to the braneworld black hole solution, the third term in $f(r)$ here is similar to the charge term in the braneworld black hole metric and hence would display identical memory effect in the presence of a gravitational wave.

\subsection{Charged Dilaton Black Holes}
\begin{figure}[H]
\begin{subfigure}{.5\textwidth}
  \centering
   \includegraphics[width=0.9\linewidth]{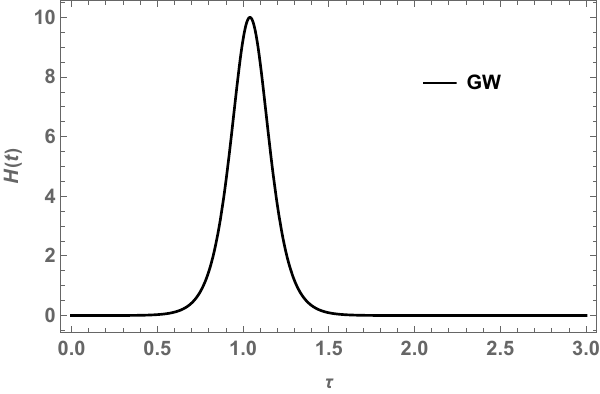}
  \caption{Gravitational wave profile}
  \label{cdgw1}
\end{subfigure}%
\begin{subfigure}{.5\textwidth}
  \centering
   \includegraphics[width=0.9\linewidth]{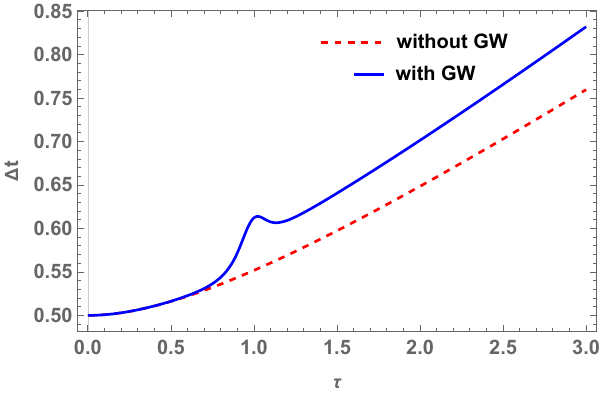}
  \caption{Deviation in $t$ co-ordinate difference}
  \label{cdgw2}
\end{subfigure}
\begin{subfigure}{.5\textwidth}
  \centering
  \includegraphics[width=0.9\linewidth]{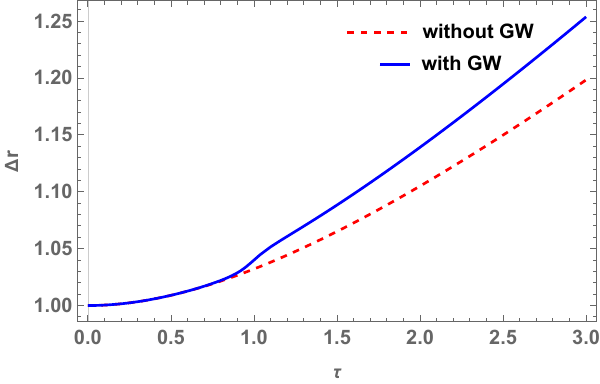}
  \caption{Deviation in $r$ co-ordinate difference}
  \label{cdgw3}
\end{subfigure}%
\begin{subfigure}{.5\textwidth}
  \centering
  \includegraphics[width=0.9\linewidth]{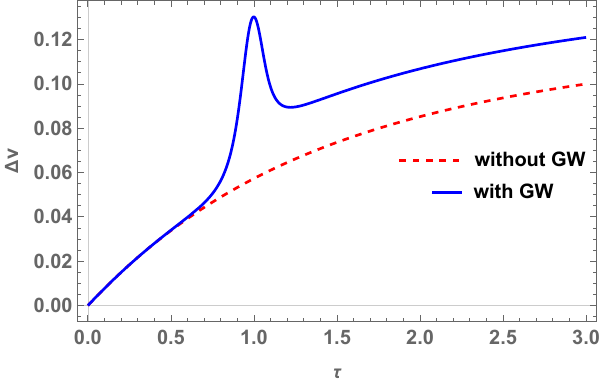}
  \caption{Deviation in velocity($v=d(\Delta r)/d\tau$) difference}
  \label{cdgw4}
\end{subfigure}
\caption{Memory effect in Charged Dilaton Black Hole solution with $D=0.01$}
\label{fig:cd}
\end{figure}

\begin{figure}[H]
\centering
\begin{subfigure}{.5\textwidth}
  \centering
   \includegraphics[width=0.9\linewidth]{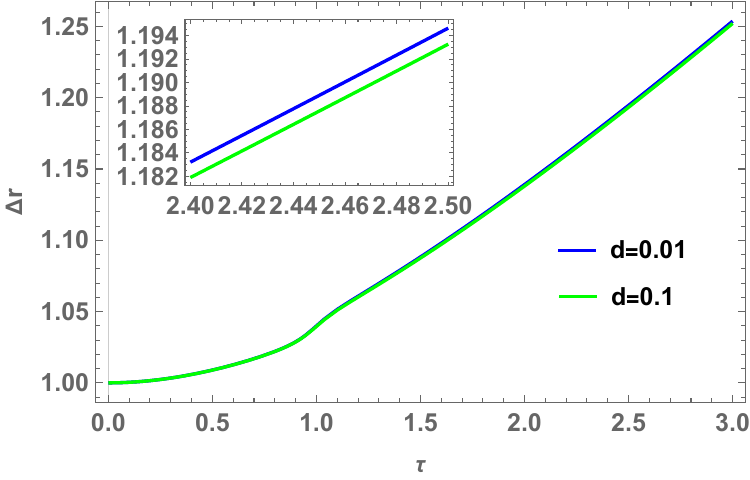}
  \caption{Comparison of $r$ co-ordinate difference}
  \label{fig:sub1}
\end{subfigure}%
\begin{subfigure}{.5\textwidth}
  \centering
  \includegraphics[width=0.9\linewidth]{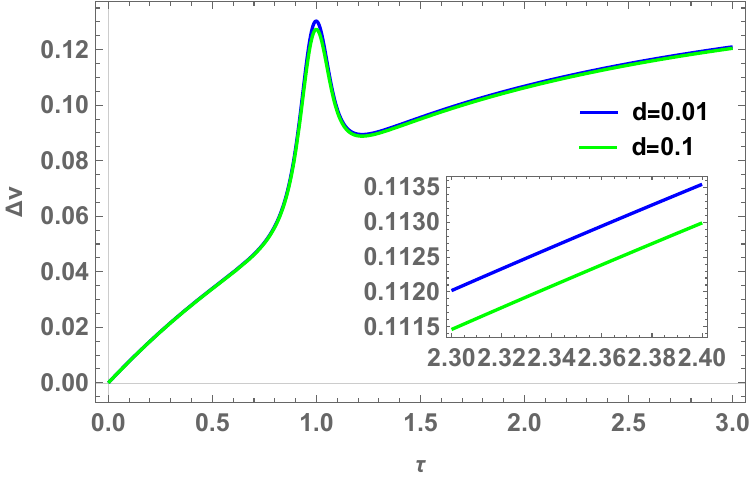}
  \caption{Comparison of velocity($v=d(\Delta r)/d\tau$) difference}
  \label{fig:sub2}
\end{subfigure}
\caption{Comparison between different values of the parameter $D$ for Charged Dilaton Black Hole solution}
\label{fig:cdcom}
\end{figure}

\noindent Here we consider the static charged black hole solution in string theory which is valid for curvature below the Planck scale and is labelled by its mass, charge and asymptotic value of the scalar field called the dilaton field $\phi$. The $4$-dimensional low energy Lagrangian that gives rise to this solution~\cite{CD} is 
\begin{equation}
    S=\int d^4x\sqrt{-g} [-R+2(\nabla \phi)^2+e^{-2\phi}F^2]
\end{equation}
\noindent where $F_{\mu \nu}$ is the Maxwell field associated with a U(1) subgroup of Spin(32)/$Z_2$ and the remaining gauge fields and anti-symmetric tensor field have been set to zero. When we extremise this Lagrangian and try to obtain a static and spherically symmetric solution with asymptotic flatness to the corresponding field equations, we get
\begin{equation}
     ds^2=-\left(1-\frac{2}{r}\right)dt^2+\frac{dr^2}{1-\frac{2}{r}}+(r^2+2Dr)d\Omega_2 ^2
\end{equation}
where  
\begin{equation}
  D=-\frac{Q^2 e^{2\phi_0}}{M}
\end{equation}
\noindent Here $\phi_0$ is the asymptotic value of the charged dilaton field and $Q$ is magnetic charge. Thus, $D$ is a constant here and we have studied memory effect for different values of $D$.
The geodesic equations in the absence of gravitational waves are:
\begin{align}
\ddot t+\frac{2}{r(r-2)}~\dot r \dot t&=0\\
\ddot r-\frac{\dot r^2}{r(r-2)}+\frac{1}{r^2}\left(1-\frac{2}{r}\right)~\dot t^2-(r+D)\left(1-\frac{2}{r}\right)~ \dot \phi^2&=0\\
\ddot \phi +2 \left(\frac{r+D}{r^2+2Dr}\right)~\dot r \dot \phi&=0
\end{align}
%The velocity normalisation condition which will set the initial values is given by
%\begin{equation}
%     -\left(1-\frac{2}{r}\right)\dot t^2+\frac{\dot r^2}{\left(1-\frac{2}{r}\right)}+(r^2+2Dr)\dot\phi^2=-1
%\end{equation}
In the presence of gravitational waves the metric looks like:
\begin{equation}
    ds^2=-\left(1-\frac{2}{r}\right)dt^2+\frac{dr^2}{\left(1-\frac{2}{r}\right)}+\left(r^2+2Dr+rH(t)\right)d\theta^2 + \left(r^2+2Dr-rH(t)\right)d\phi^2
\end{equation}
And the corresponding geodesic equations look like
\begin{align}
\ddot t+\frac{2}{r(r-2)}~\dot r \dot t-\left(\frac{r^2 H'(t)}{2r-4}\right)~\dot \phi^2&=0\\
\ddot r-\frac{\dot r^2}{r(r-2)}+\frac{\dot t^2}{r^2}\left(1-\frac{2}{r}\right)-\left(\frac{2r+2D-H(t)}{2}\right)\left(1-\frac{2}{r}\right)~\phi^2&=0\\
\ddot \phi+\left(\frac{2r\dot r+2D\dot r-H(t)\dot r-H'(t)\dot tr}{r^2+2Dr-rH(t)}\right)~\dot \phi &=0
\end{align}
\noindent The displacement and velocity memory effect has been depicted in figure \ref{fig:cd}. 
A comparison for different values of the parameter $D$ is shown in figure \ref{fig:cdcom}. 

\subsection{Boulware-Deser black hole solution}
\noindent The spherically symmetric static solution of Einstein-Gauss-Bonnet theory was obtained by Boulware and Deser in \cite{Deserb} and a simpler form of the metric is given in \cite{BDpaper} 
\begin{equation}
    ds^2=-f(r)dt^2+\frac{dr^2}{f(r)}+r^2d\psi^2+r^2\sin^2\psi d\theta^2+r^2\sin^2\psi\sin^2\theta d\phi^2
\end{equation}
where
\begin{equation}
    f(r)=1+\frac{r^2}{4\alpha}\left(1+\sigma \sqrt{1+\frac{16\alpha M}{r^4}+\frac{4\alpha\Lambda}{3}}\right)
\end{equation}
\noindent Here, $\sigma^2=1$ and $\Lambda$ is the cosmological constant. This is the most general spherically symmetric solution to the Einstein-Gauss-Bonnet theory, on the condition that the metric is smooth everywhere. For $\alpha>0$ and $\sigma=-1$, this solution represents a black hole whose horizon is located at $r_+=\sqrt{2(M-\alpha)}$, given that $\Lambda=0$. However, for $\alpha>0, ~M>0$ and $\sigma=+1$, this solution has a naked singularity at $r=0$. In this paper, we study the former case of a black hole solution. In that case, $f(r)$ takes the form (putting $M=1$),
\begin{equation}
    f(r)=1+\frac{r^2}{4\alpha}\left(1-\sqrt{1+\frac{16\alpha}{r^4}}\right)
\end{equation}
Let us consider that $\theta$ is fixed at $\pi/2$. Our metric then becomes
\begin{equation}
    ds^2=-f(r)dt^2+\frac{dr^2}{f(r)}+r^2d\psi^2+r^2\sin^2\psi d\phi^2
\end{equation}
In the absence of a gravitational wave, the geodesic equations look like:
\begin{align}
    \ddot t+\frac{r \left(2-2 \sqrt{\frac{16 \alpha }{r^4}+1}\right)}{\sqrt{\frac{16 \alpha }{r^4}+1} \left(r^2 \left(\sqrt{\frac{16 \alpha }{r^4}+1}-1\right)-4 \alpha \right)}~\dot r \dot t&=0\\
    \ddot r-\frac{\dot r^2}{2} \frac{r \left(2-2 \sqrt{\frac{16 \alpha }{r^4}+1}\right)}{\sqrt{\frac{16 \alpha }{r^4}+1} \left(r^2 \left(\sqrt{\frac{16 \alpha }{r^4}+1}-1\right)-4 \alpha \right)}+\frac{r \left(\sqrt{\frac{16 \alpha }{r^4}+1}-1\right) \left(r^2 \left(\sqrt{\frac{16 \alpha }{r^4}+1}-1\right)-4 \alpha \right)}{16 \alpha ^2 \sqrt{\frac{16 \alpha }{r^4}+1}}~\dot t^2- \nonumber \\ r\left(1+\frac{r^2}{4\alpha}\left(1-\sqrt{1+\frac{16\alpha M}{r^4}}\right)\right)\left(\dot \psi^2+\sin^2\psi~\dot \phi^2\right)&=0\\
    \ddot \psi+\frac{2}{r}~\dot \psi \dot r-\sin\psi \cos\psi~\dot\phi^2&=0\\
    \ddot \phi+\frac{2}{r}~\dot r\dot \phi+\frac{2\cos\psi}{\sin\psi}~\dot \phi \dot \psi&=0
\end{align}
\noindent Now in the presence of a gravitational wave, the metric is given by,
\begin{equation}
      ds^2=-\left(1+\frac{r^2}{4\alpha}\left(1-\sqrt{1+\frac{16\alpha }{r^4}}\right)\right)dt^2+\frac{dr^2}{1+\frac{r^2}{4\alpha}\left(1-\sqrt{1+\frac{16\alpha }{r^4}}\right)}+\left(r^2+rH(t)\right)\psi^2+\left(r^2-rH(t)\right)\sin^2\psi d\phi^2
\end{equation}
The geodesic equations then look like
\begin{align}
    \ddot t+\frac{r \left(2-2 \sqrt{\frac{16 \alpha }{r^4}+1}\right)}{\sqrt{\frac{16 \alpha }{r^4}+1} \left(r^2 \left(\sqrt{\frac{16 \alpha }{r^4}+1}-1\right)-4 \alpha \right)}~\dot r \dot t+\frac{rH'(t)}{2}\frac{4 \alpha }{r^2 \left(\sqrt{\frac{16 \alpha }{r^4}+1}-1\right)-4 \alpha }\left(\dot \psi^2-\sin^2\psi~\dot \phi^2\right)&=0\\
    \ddot r-\frac{\dot r^2}{2}\frac{r \left(2-2 \sqrt{\frac{16 \alpha }{r^4}+1}\right)}{\sqrt{\frac{16 \alpha }{r^4}+1} \left(r^2 \left(\sqrt{\frac{16 \alpha }{r^4}+1}-1\right)-4 \alpha \right)}+\frac{r \left(\sqrt{\frac{16 \alpha }{r^4}+1}-1\right) \left(r^2 \left(\sqrt{\frac{16 \alpha }{r^4}+1}-1\right)-4 \alpha \right)}{16 \alpha ^2 \sqrt{\frac{16 \alpha }{r^4}+1}}~\dot t^2 \nonumber \\ -\left(1+\frac{r^2}{4\alpha}\left(1-\sqrt{1+\frac{16\alpha M}{r^4}}\right)\right)\left(\left(2r+H(t)\right)\frac{\dot \psi^2}{2}+\sin^2\psi\left(2r-H(t)\right)\frac{\dot \phi^2}{2}\right)&=0\\
    \ddot \psi+ \left(\frac{2r\dot r+\dot rH(t)+rH'(t)\dot t}{r^2+rH(t)}\right)\dot \psi-\sin\psi \cos\psi \left(\frac{r^2-rH(t)}{r^2+rH(t)}\right)\dot \phi^2&=0\\
    \ddot \phi+2\frac{\cos\psi}{\sin\psi}~\dot \phi \dot \psi+\left(\frac{2r\dot r-\dot rH(t)-rH'(t)\dot t}{r^2-rH(t)}\right)\dot \phi&=0
\end{align}

\begin{figure}[H]
\centering
\begin{subfigure}{.5\textwidth}
  \centering
  \includegraphics[width=0.9\linewidth]{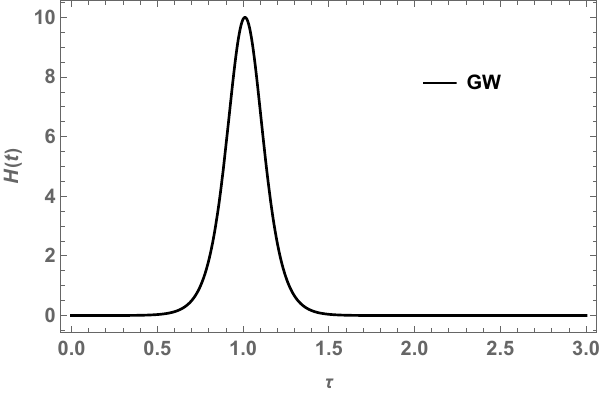}
  \caption{Gravitational wave profile }
  \label{bd1}
\end{subfigure}%
\begin{subfigure}{.5\textwidth}
  \centering
   \includegraphics[width=0.9\linewidth]{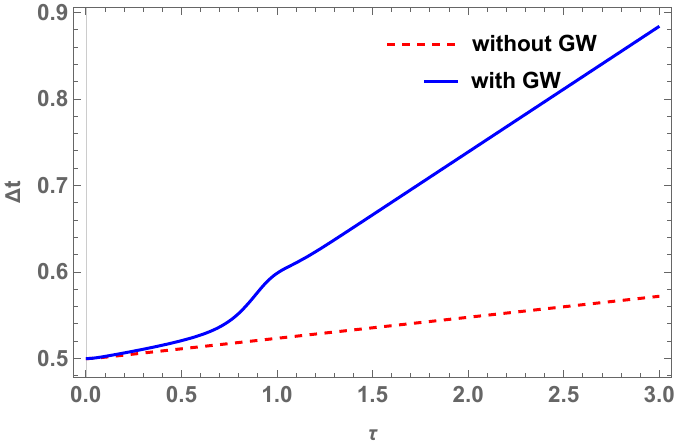}
  \caption{Deviation in $t$ coordinate difference}
  \label{bd2}
\end{subfigure}%

\begin{subfigure}{.5\textwidth}
  \centering
  \includegraphics[width=0.9\linewidth]{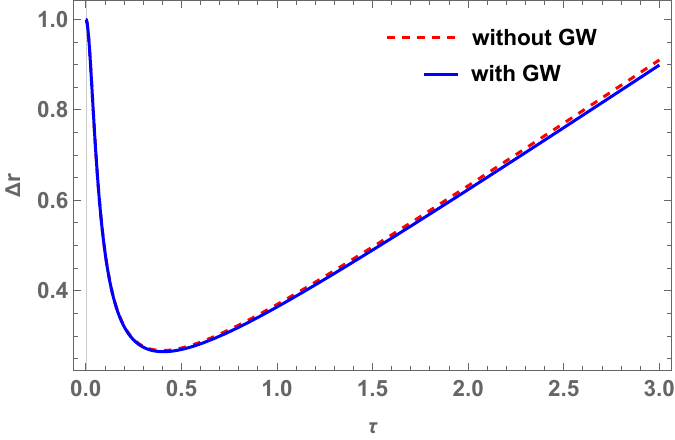}
  \caption{Deviation in $r$ coordinate difference}
  \label{bd3}
\end{subfigure}%
\begin{subfigure}{.5\textwidth}
  \centering
  \includegraphics[width=0.9\linewidth]{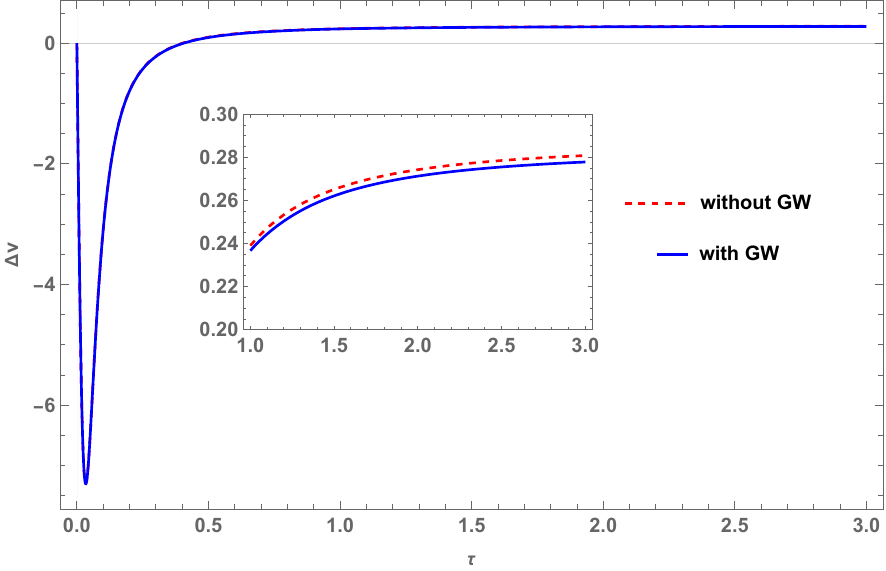}
  \caption{Deviation in velocity($v=d(\Delta r)/d\tau$) difference}
  \label{bd4}
\end{subfigure}%

\caption{Memory effect in Boulware-Deser solution with $\alpha=0.01$}
\label{fig:bd}
\end{figure}

\noindent The displacement and velocity memory effect for the Boulware-Deser solution is depicted in figure \ref{fig:bd}. From these figures one can see that the deviation of $\Delta r$ with $\tau$ is extremely small and from the general behaviour in previous plots we can predict that the corresponding velocity variations would be even smaller in scale. Hence in the velocity memory plot in figure \ref{bd4}, the memory effect is not quite visible because the deviation is much smaller than the vertical scale of the plot. However, when we compared the exact numerical values, we did notice some deviation as can be seen from the magnified region as shown in the inset of figure \ref{bd4}.

\section{Memory effect in Static and Spherically symmetric Wormhole solutions} \label{sec4}
\noindent Wormholes are solutions of Einstein's field equations that are characterised by the absence of an event horizon and the presence of a throat connecting two distant regions in spacetime. These are usually unstable structures that require exotic matter to sustain and hence are till-date hypothetical in nature. However they still serve as good toy models to study regions of extreme gravity. Here we consider some static and spherically symmetric wormhole solutions and explore the GW memory by analysing the geodesic evolution as has been done for Schwarzschild case.
\subsection{Damour Solodukhin Wormhole}
\noindent  Let us consider the simplest spherically symmetric wormhole solution which was given by Damour and Solodukhin \cite{damsol} whose metric is given by,
\begin{equation}
 ds^2=-\Big(f(r)+\lambda^2\Big)dt^2 + \frac{dr^2}{f(r)} + r^2d\Omega_2 ^2   
\end{equation}
here, $ f(r) = 1-2/r$. This is a wormhole solution as is evident from the fact that there is no event horizon because the null horizon that we get from $g^{rr}$ does not coincide with the killing horizon. However, it becomes an event horizon when $\lambda=0$ in which case we get back the Schwarzschild metric. For non-zero values of $\lambda$ we get the usual wormhole structure, i.e. absence of an event horizon and a throat region at $r = 2M$. This is an example of a Lorentzian wormhole \cite{lorentzianwh}. The Damour-Solodukhin metric also exhibits bizarre features, for example, the $G_{tt} $ component vanishes which implies that matter with vanishing energy density is required to sustain such a structure~\cite{echoeskerr}.
Let us consider some substitutions in this metric as $t$ here does not correspond to asymptotic observer, therefore, performing $t \rightarrow 1/\sqrt{1+\lambda^2}$  and 
$M \rightarrow M(1+\lambda^2)$. The metric now becomes, 
\begin{equation}
      ds^2=-\left(1-\frac{2M}{r} \right)dt^2 + \frac{dr^2}{1- \frac{2M(1+\lambda^2)}{r}} + r^2d\Omega_2 ^2   
\end{equation}
%\noindent The velocity normalisation condition\ref{vel norm} helps us determine the initial conditions.
%\begin{equation}
%    -\left(1-\frac{2M}{r} \right)~\dot t^2 + \frac{1}{1- \frac{2M(1+\lambda^2)}{r}}\dot r^2+r^2 \dot \phi^2=-1
%\end{equation}
The equations of motion in the absence of a gravitational wave would look like,
\begin{align}
\ddot t +\frac{2}{r(r-2)} ~\dot r\dot{t}&= 0\\
\ddot r -\frac{\dot r^2}{r} \left(\frac{1+\lambda^2}{r-2(1+\lambda^2)}\right)+\left(1-\frac{2(1+\lambda^2)}{r}\right)\left(\frac{\dot t^2}{r^2}-r\dot \phi^2\right) &= 0\\ 
\ddot \phi + \frac{2}{r} ~\dot r\dot \phi&= 0  
\end{align}
\noindent Using the above equations of motion, we obtained two solutions for two geodesics, each with same initial conditions for $\phi(0)$, $\dot \phi(0), ~\dot r(0)$ and $\dot t(0)$ but differing in the initial values $r(0)$ and $t(0)$.
We then compute the $\Delta r$, $\Delta t$ and $\Delta v$ values which are the difference between the respective coordinates in each geodesic solution and plot them.
Now, to see the effect of a passing gravitational wave in this spacetime, we must modify the above metric. We can do this by keeping in mind that gravitational waves are described in TT gauge and thus (considering zero cross-polarisation) we modify the $g_{\theta\theta}$ and $g_{\phi\phi}$ components.\\
\noindent In the presence of gravitational wave, the metric becomes,
\begin{equation}
    ds^2=-\left(1-\frac{2M}{r} \right)dt^2+\frac{dr^2}{{1- \frac{2M(1+\lambda^2)}{r}}}+\Big(r^2+rH(t) \Big)d\theta^2
+\Big(r^2-rH(t)\Big)\sin^2\theta d\phi^2
\end{equation}
And the corresponding geodesic equations look like:
\begin{align}
\ddot t  +\frac{2}{r(r-2)} ~\dot r\dot{t}-\frac{r^2H'(t)}{2(r-2)}~\dot \phi^2 &= 0\\
\ddot{r}-\frac{\dot r^2}{r} \left(\frac{1+\lambda^2}{r-2(1+\lambda^2)}\right)+\left(1-\frac{2(1+\lambda^2)}{r}\right)\left(\frac{\dot t^2}{r^2}  +\frac{(H(t)-2r)}{2}\dot \phi^2\right) &=0\\
\ddot{\phi}+\left( \frac{2r\dot{r}-\dot{r}H(t)-rH'(t)\dot{t}}{r^2-rH(t)} \right ) \dot{\phi}&=0
\end{align}

\begin{figure}[H]
\centering
\begin{subfigure}{.5\textwidth}
  \centering
   \includegraphics[width=0.9\linewidth]{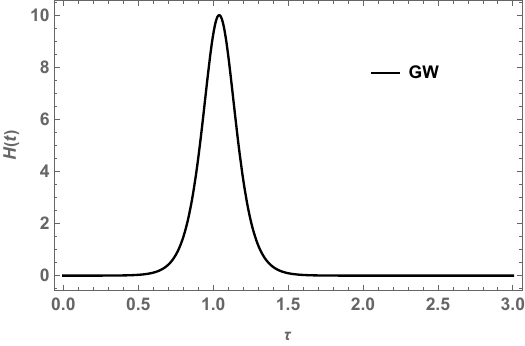}
  \caption{Gravitational wave profile}
  \label{ds1}
\end{subfigure}%
\begin{subfigure}{.5\textwidth}
  \centering
   \includegraphics[width=0.9\linewidth]{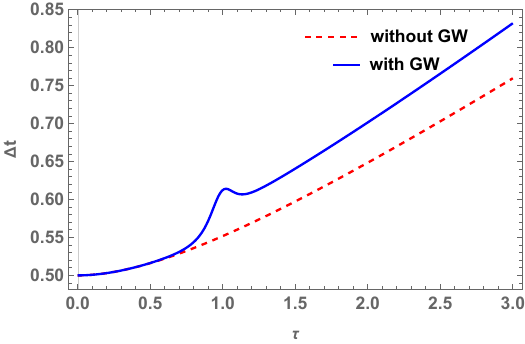}
  \caption{Deviation in $t$ co-ordinate difference}
  \label{ds22}
\end{subfigure}%

\begin{subfigure}{.5\textwidth}
  \centering
  \includegraphics[width=0.9\linewidth]{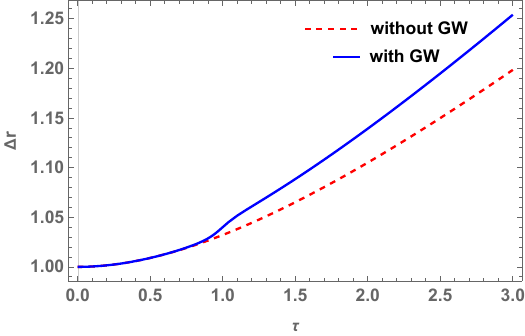}
  \caption{Deviation in $r$ co-ordinate difference}
  \label{ds3}
\end{subfigure}%
\begin{subfigure}{.5\textwidth}
  \centering
  \includegraphics[width=0.9\linewidth]{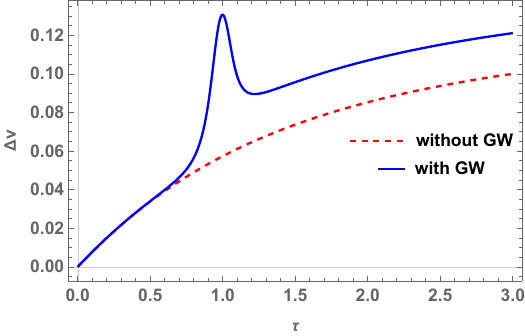}
  \caption{Deviation in velocity($v=d(\Delta r)/d\tau$) difference}
  \label{ds4}
\end{subfigure}%
\caption{Memory effect in Damour-Solodukhin wormhole with $\lambda=0.01$}
\label{fig:ds}
\end{figure}

\noindent The displacement and velocity memory effects for the Damour-Solodukhin wormhole solution have been shown in the figure \ref{fig:ds}.

%\newpage

\begin{figure}[hbt!]
\centering
\begin{subfigure}{.5\textwidth}
  \centering
  \includegraphics[width=0.9\linewidth]{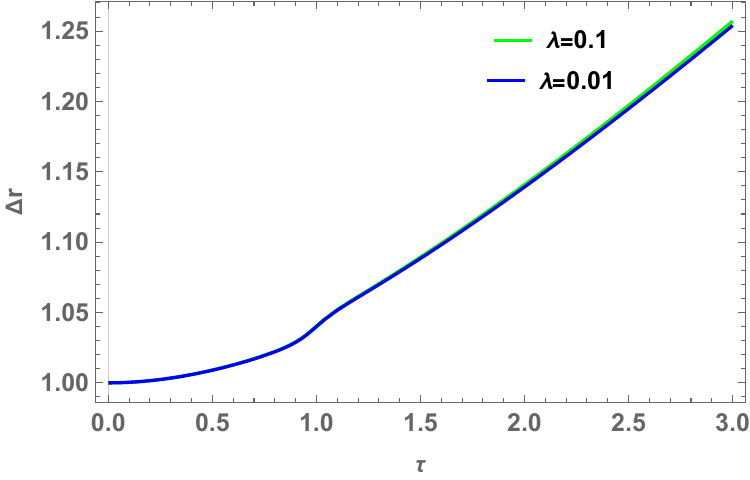}
  \caption{Comparison of $r$ co-ordinate difference}
  \label{fig:sub1}
\end{subfigure}%
\begin{subfigure}{.5\textwidth}
  \centering
  \includegraphics[width=0.9\linewidth]{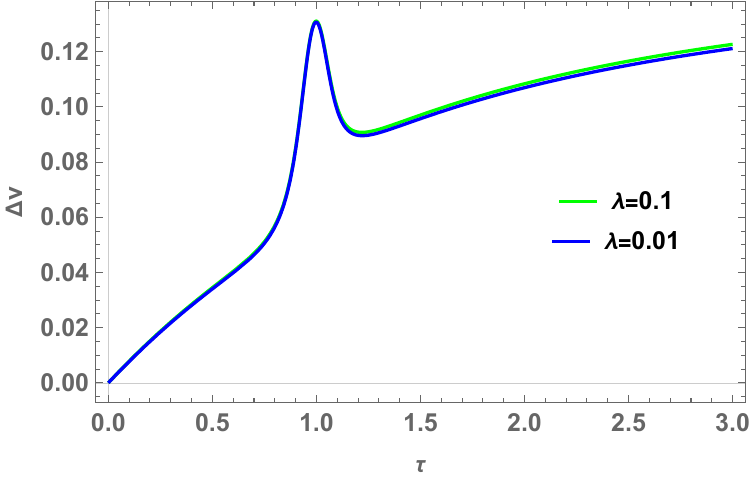}
  \caption{Comparison of velocity ($v=d(\Delta r)/d\tau$) difference }
  \label{fig:sub2}
\end{subfigure}
\caption{Comparison between different values of $\lambda$ for Damour-Solodukhin wormhole}
\label{fig:dscom}
\end{figure}

\noindent Since the Damour-Solodukhin metric depends on the wormhole hair $\lambda$, we would like to see the how the memory effects depend on the wormhole hair $\lambda$.  We depict the effect in figure \ref{fig:dscom} for two different values $\lambda=0.01$ and $\lambda=0.1$.
%to show that memory effect is not a result of any specific value of the metric parameter. Note that the parameter $\lambda$ must take values less than one for a valid solution.

\subsection{Wormhole solution in Kalb-Ramond Theory}
\noindent The Einstein-Kalb-Ramond theory is essentially a scalar coupled theory which involves a term $H_{\mu\nu\lambda}$, which is the source term for the gauge field, that is antisymmetric in three indices and hence is interpreted as the torsion factor that arises in covariant derivative of a tensor when the indices of the Christoffel symbol are antisymmetric. 
The action~\cite{ssgpaper} for this gauge-invariant theory is
\begin{equation}
    S=\int d^4x \sqrt{-g}\left(\frac{R(g)}{\kappa}-\frac{1}{12}H_{\mu\nu\lambda}H^{\mu\nu\lambda}\right )
\end{equation}
$R(g)$ is the Ricci scalar curvature and $\kappa\sim({\rm Planck~mass})^{-2}$ is the gravitational coupling constant.
Here we consider the Einstein-Kalb-Ramond theory in $4$-dimensions where the simplest static and spherically symmetric solution in this theory looks like:
\begin{equation}
    ds^2=-e^{\nu(r,t)}dt^2+e^{\lambda(r,t)}dr^2+r^2d\theta^2+r^2\sin^2\theta d\phi^2
\end{equation}
\noindent Again it is evident from the metric why this is a wormhole solution and not a black hole solution because there is no event horizon. Using the following substitutions gives us a static and spherically symmetric wormhole for a real Kalb-Ramond field \cite{ssgpaper}, 
\begin{equation}
     e^{\nu}=1, ~~~~~~~~~~~~~~~~~e^{-\lambda}=1-\frac{b}{r^2}
\end{equation} 
 where $b$ is a positive constant and captures the information of the Kalb Ramond field.\\
The metric then becomes 
\begin{equation}
    ds^2=-dt^2+\frac{dr^2}{1-\frac{b}{r^2}} + r^2d\Omega_2 ^2 \label{krm}
\end{equation}
%Considering $\theta=\pi/2$, the velocity normalisation condition for time-like geodesic is $g_{\mu\nu}u^{\mu}u^{\nu}=-1$
%\begin{equation}
%    -\dot t^2 + \frac{1}{1-\frac{b}{r^2}}\dot r^2+r^2\dot \phi^2=-1
%\end{equation}
The corresponding geodesic equations are:
\begin{align}
    \ddot t &=0 \\
    \ddot r-\frac{\dot r^2}{r(r^2-b)}-\left(r-\frac{b}{r}\right) ~\dot \phi^2&=0\\
    \ddot \phi + \frac{2}{r}~ \dot r \dot \phi&=0
\end{align}

\begin{figure}[H]
\begin{subfigure}{.5\textwidth}
\centering
 \includegraphics[width=0.9\linewidth]{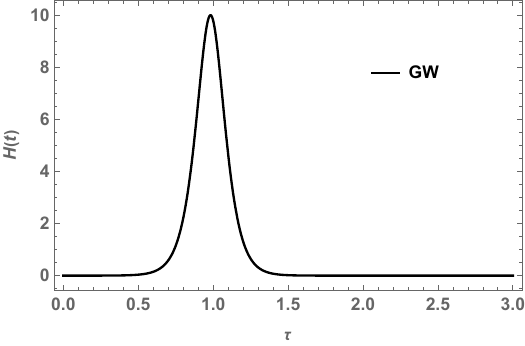}
  \caption{Gravitational wave profile}
  \label{kr1}
\end{subfigure}%
\begin{subfigure}{.5\textwidth}
  \centering
   \includegraphics[width=0.9\linewidth]{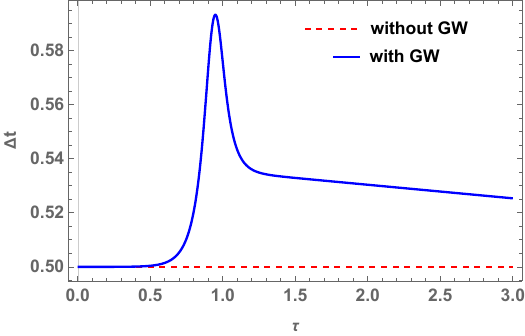}
  \caption{Deviation in $t$ co-ordinate difference}
  \label{kr2}
\end{subfigure}%

\begin{subfigure}{.5\textwidth}
  \centering
  \includegraphics[width=0.9\linewidth]{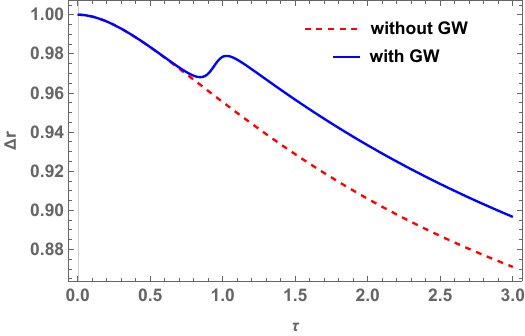}
  \caption{Deviation in $r$ co-ordinate difference}
  \label{kr3}
\end{subfigure}%
\begin{subfigure}{.5\textwidth}
  \centering
  \includegraphics[width=0.9\linewidth]{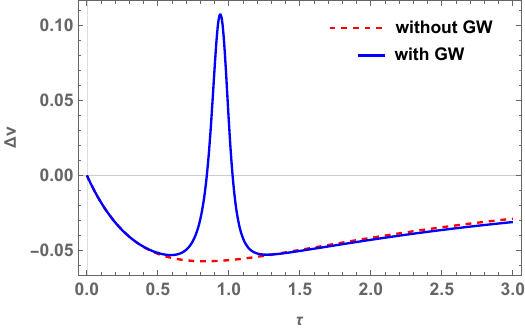}
  \caption{Deviation in velocity($v=d(\Delta r)/d\tau$) difference}
  \label{kr4}
\end{subfigure}%
\caption{Memory effect in Static and Spherically symmetric solution of Kalb Ramond field with $b=0.1$}
\label{fig:kr}
\end{figure}

\begin{figure}[hbt!]
\begin{subfigure}{.5\textwidth}
  \centering
  \includegraphics[width=0.9\linewidth]{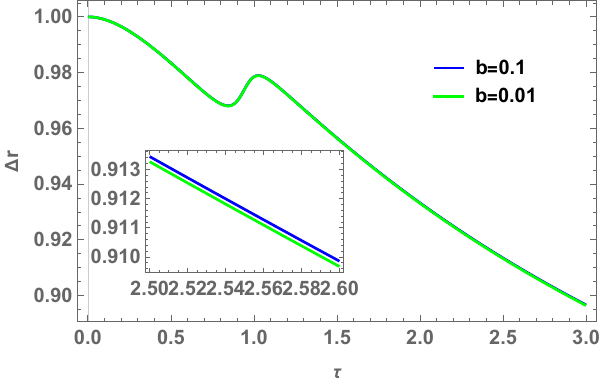}
  \caption{Comparison of $r$ co-ordinate difference}
  \label{fig:sub1}
\end{subfigure}%
\begin{subfigure}{.5\textwidth}
  \centering
  \includegraphics[width=0.9\linewidth]{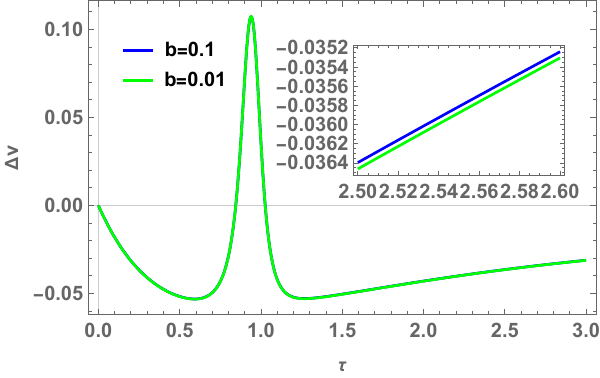}
  \caption{Comparison of velocity ($v=d(\Delta r)/d\tau$) difference }
  \label{fig:sub2}
\end{subfigure}
\caption{Comparison between different values of the parameter $b$ in Static and Spherically symmetric solution of Kalb-Ramond field}
\label{fig:krcom}
\end{figure}

\noindent Now, in the presence of a gravitational wave of the form \ref{wave}, this metric would look like :
\begin{equation}
    ds^2=-dt^2+\frac{dr^2}{1-\frac{b}{r^2}} +\Big(r^2+rH(t) \Big)d\theta^2
+\Big(r^2-rH(t)\Big)\sin^2\theta d\phi^2
\end{equation}
\noindent And the corresponding geodesic equations are
 \begin{align}
     \ddot t-\frac{rH'(t)}{2}~\dot \phi^2&=0\\
    \ddot r+ \frac{b\dot r^2}{r(r^2-b)} - \left(\frac{2r-H(t)}{2}\right)\left(1-\frac{b}{r^2}\right)~\dot \phi^2&=0\\
    \ddot \phi+\left( \frac{2 r \dot r-H(t) \dot r-r \dot t H'(t)}{r^2-r H(t)}\right)~\dot \phi &=0
 \end{align}
The displacement and velocity memory effects for the Kalb-Ramond wormhole are depicted in figure \ref{fig:kr}.
\noindent The metric \ref{krm} depends on the wormhole hair $b$. We show in figure \ref{fig:krcom} how the memory effect depend on the wormhole hair $b$ for two different values of $b$ from $O(10^{-1})$ to $O(1)$ and thus notice a difference.

\noindent We again try to demonstrate memory effect in a static and spherically symmetric wormhole solution of the Kalb-Ramond theory but this time we consider a more general expression given in \cite{kr}, which is of the form \ref{sphm}, with $f(r)$ and $g(r)$ are given by,
\begin{align}
    f(r)&= 1-\frac{2}{r}-\frac{b}{3r^3}\\
    g(r)&=1-\frac{2}{r}-\frac{b}{r^2}
\end{align}
%\noindent The velocity normalisation condition\ref{vel norm} for time-like geodesic looks like:
%\begin{equation}
%    -\left( 1-\frac{2}{r}-\frac{b}{3r^3}\right)\dot t^2 + \frac{\dot r^2}{1-\frac{2}{r}-\frac{b}{r^2}}+r^2\dot \phi^2=-1
%\end{equation}
\noindent This equation helps us set the initial conditions for the geodesic evolution. 
\noindent For the given metric, in the absence of a gravitational wave, the geodesic equations are:
\begin{align}
\ddot t+\frac{3}{r}\left(\frac{2r^2+b}{3r^3-6r^2-b}\right)~\dot r \dot t&=0\\
\ddot r-\left(\frac{r+b}{r^2-2r-b}\right)\frac{\dot r^2}{r} + \frac{(2r^2+b)(r^2-2r-b)}{2r^6}~\dot t^2 - \left(r-2-\frac{b}{r}\right)~\dot \phi^2 &=0\\
\ddot \phi+\frac{2 }{r}~\dot r \dot \phi&=0
\end{align}
\noindent In presence of a gravitational wave of the form \ref{wave} in TT gauge, the metric becomes
\begin{equation}
    ds^2=-\left(1-\frac{2}{r}-\frac{b}{3r^3}\right)dt^2 + \frac{dr^2}{1-\frac{2}{r}-\frac{b}{r^2}} +\Big(r^2+rH(t)\Big)d\theta^2 + \Big(r^2-rH(t)\Big)\sin^2\theta d\phi^2
\end{equation}
The corresponding equations of motion are
\begin{align}
\ddot t-\frac{3r^4 H'(t)}{2(3r^3-6r^2-b)}~\dot \phi ^2+\frac{3}{r}\left(\frac{2r^2+b}{3r^3-6r^2-b}\right)~\dot r \dot t&=0\\
\ddot r-\left(\frac{r+b}{r^2-2r-b}\right)\frac{\dot r^2}{r} + \frac{(2r^2+b)(r^2-2r-b)}{2r^6}~\dot t^2+\left(\frac{H(t)-2r}{2}\right)\left(1-\frac{2}{r}-\frac{b}{r^2}\right)~\dot \phi^2&=0\\
\ddot \phi +\left(\frac{2r \dot r-\dot r H(t)-rH'(t)\dot t}{r^2-rH(t)}\right )~\dot \phi &=0
\end{align}

\begin{figure}[H]
\begin{subfigure}{.5\textwidth}
  \centering
   \includegraphics[width=0.9\linewidth]{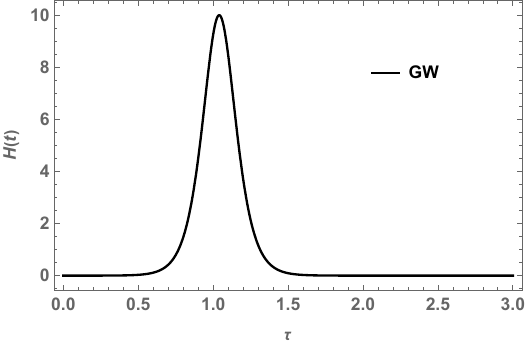}
  \caption{Gravitational wave profile}
  \label{fig:sub1}
\end{subfigure}
\begin{subfigure}{.5\textwidth}
  \centering
   \includegraphics[width=0.9\linewidth]{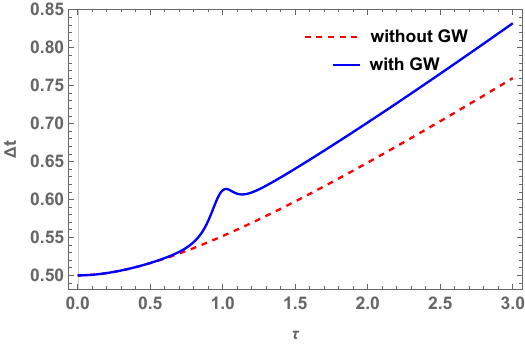}
  \caption{Deviation in $t$ co-ordinate difference}
  \label{fig:sub1}
\end{subfigure}
\begin{subfigure}{.5\textwidth}
  \centering
  \includegraphics[width=0.9\linewidth]{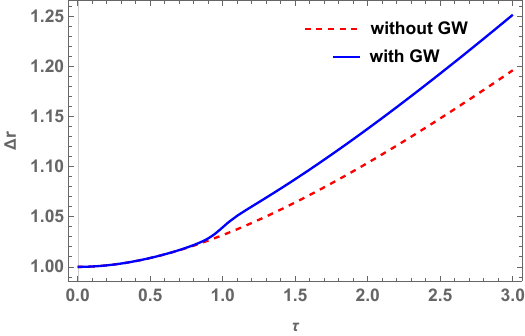}
  \caption{Deviation in $r$ co-ordinate difference}
  \label{fig:sub2}
\end{subfigure}
\begin{subfigure}{.5\textwidth}
  \centering
  \includegraphics[width=0.9\linewidth]{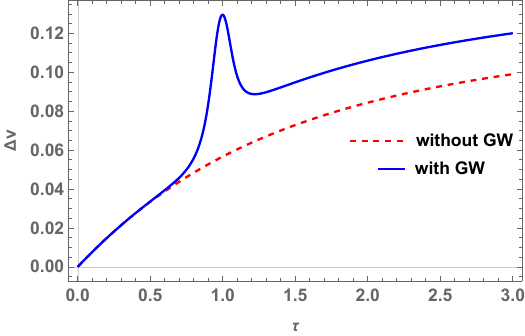}
  \caption{Deviation in velocity($v=d(\Delta r)/d\tau$) difference}
  \label{fig:sub1}
\end{subfigure}
\caption{Memory effect in static and spherically symmetric solution in Kalb-Ramond theory with $b=0.1$}
\label{fig:krmod}
\end{figure}

\begin{figure}[H]
\begin{subfigure}{.5\textwidth}
  \centering
  \includegraphics[width=0.9\linewidth]{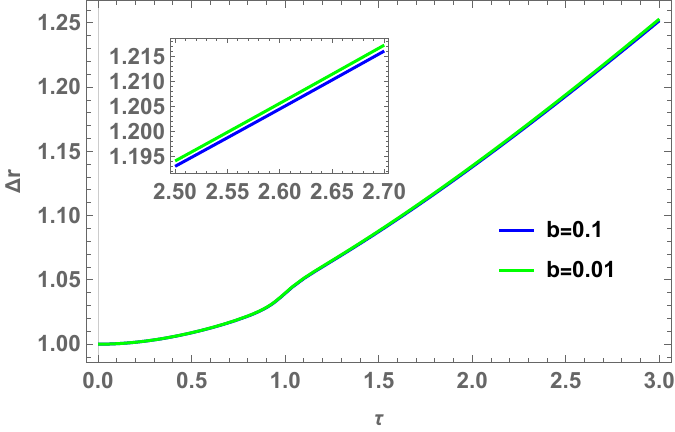}
  \caption{Comparison of $r$ co-ordinate difference}
  \label{fig:sub1}
\end{subfigure}%
\begin{subfigure}{.5\textwidth}
  \centering
  \includegraphics[width=0.9\linewidth]{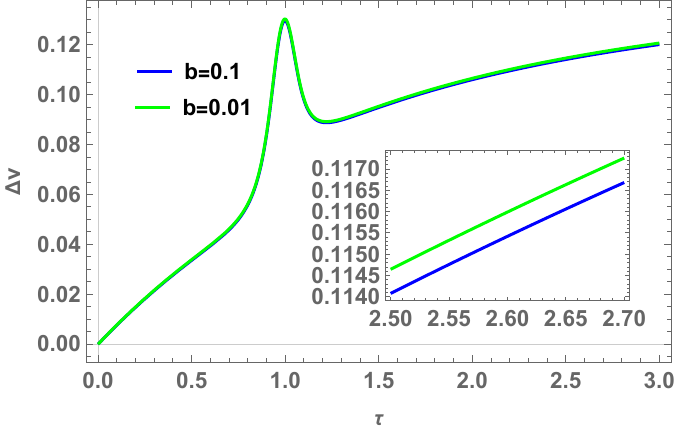}
  \caption{Comparison of velocity ($v=d(\Delta r)/d\tau$) difference }
  \label{fig:sub2}
\end{subfigure}
\caption{Comparison between different choices for the parameter $b$ in static and spherically symmetric solution of the Kalb-Ramond theory}
\label{fig:krmodcom}
\end{figure}

\noindent The displacement and velocity memory effects for this general form of Kalb-Ramond solution are depicted in figure \ref{fig:krmod}.
\noindent The dependence of the memory effect on the wormhole hair $b$ is shown in figure \ref{fig:krmodcom}.

\subsection{Braneworld wormholes}
\noindent The braneworld theory~\cite{braneworld} is a higher dimensional theory of spacetime which says that all matter in our universe exists on a four dimensional brane. The length between two such branes may be dynamic in nature and is filled with the five dimensional bulk. The advantage of working with wormholes in the braneworld scenario is that most models of wormholes require exotic matter to sustain such structures which gives rise to questions regarding their stability and existence but in this case we can avoid dealing with exotic matter because of the presence of a higher dimension. The exotic matter is an attribute of the five dimensional bulk and since we live in four dimensions, we can work around it. A Braneworld Wormhole connects spacetime regions on the same brane. A detailed analysis of GW memory in this background has been provided in \cite{SC4} using Bondi-Sachs coordinates. 
Here we write down the metric as follows
\begin{equation}
    ds^2=-\left(\alpha+\lambda \sqrt{1-\frac{2}{r}}\right)^2 dt^2 + \frac{dr^2}{1-\frac{2}{r}} + r^2d\theta^2 + r^2\sin^2\theta d\phi^2
\end{equation}
where $\alpha$ and $\lambda$ are taken to be real and positive to avoid the formation of a naked singularity. However, the above metric is not asymptotically flat. Hence we redefine the time coordinate as $t \to t/(\alpha+t)$ and write the metric in terms of a new parameter $p=\alpha/\lambda$
\begin{equation}
    ds^2=-\left(\frac{p+\sqrt{1-\frac{2}{r}}}{p+1}\right)^2dt^2 + \frac{dr^2}{1-\frac{2}{r}} + r^2d\theta^2 + r^2\sin^2\theta d\phi^2
\end{equation}
\noindent Again note that this is not a black hole but a wormhole solution because for non-zero values of the parameter $p$, the $r=2M$ surface is a null surface but is not the killing horizon for the killing vector $\xi_t^\mu=\left(\partial/\partial t\right)^\mu$ and hence there is no event horizon. However, the $p=0$ limit gives back the original Schwarzschild black hole metric. The corresponding geodesic equations are:
\begin{align}
\ddot t+\frac{2 \dot r \dot t}{r \left(p \sqrt{1-\frac{2}{r}} r+r-2\right)}&=0\\
\ddot r+\frac{\sqrt{1-\frac{2}{r}} \left(p+\sqrt{1-\frac{2}{r}}\right) \dot t^2}{(p+1)^4 r^2}+\frac{\dot r^2}{2 r-r^2}-(r-2) \dot \phi^2&=0\\
\ddot \phi+\frac{2 }{r}~\dot r \dot \phi&=0
\end{align}
%We use the velocity normalisation condition for time-like geodesics to set the initial conditions:
%\begin{equation}
%    -\left(\frac{p+\sqrt{1-\frac{2}{r}}}{p+1}\right)\dot t^2+\frac{\dot r^2}{1-\frac{2}{r}}+r^2\dot \phi^2=-1
%\end{equation}
Whereas in the presence of a gravitational wave in TT gauge with a pulse profile, the metric looks like
\begin{equation}
    ds^2=-\left(\frac{p+\sqrt{1-\frac{2}{r}}}{p+1}\right)^2dt^2 + \frac{dr^2}{1-\frac{2}{r}} + \Big(r^2+rH(t)\Big)d\theta^2 + \Big(r^2-rH(t)\Big)\sin^2\theta d\phi^2
\end{equation}
And the corresponding geodesic equations are:
\begin{align}
\ddot t+\frac{2 ~\dot r \dot t}{r \left(p \sqrt{1-\frac{2}{r}} r+r-2\right)}-\frac{ r  H'(t)(p+1)^2}{2 \left(p+\sqrt{1-\frac{2}{r}}\right)^2}~\dot \phi^2&=0\\
\ddot r+\frac{r-2}{r^3(1+p)^2}\left(1+\frac{p}{\sqrt{1-\frac{2}{r}}}\right)~\dot t^2- \frac{\dot r^2}{r(r-2)}+\left(\frac{ H(t)-r}{2}\right)\left(1-\frac{2}{r}\right)~\dot \phi^2&=0 \\
\ddot \phi + \left(\frac{2r \dot r-\dot r H(t)-rH'(t)\dot t}{r^2-rH(t)}\right )\dot \phi&=0
\end{align}

\begin{figure}[H]
\begin{subfigure}{.5\textwidth}
  \centering
   \includegraphics[width=0.9\linewidth]{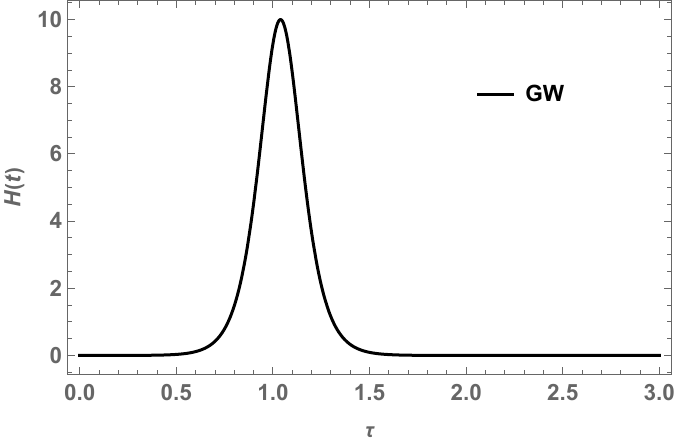}
  \caption{Gravitational wave profile}
  \label{fig:sub1}
\end{subfigure}%
\begin{subfigure}{.5\textwidth}
  \centering
   \includegraphics[width=0.9\linewidth]{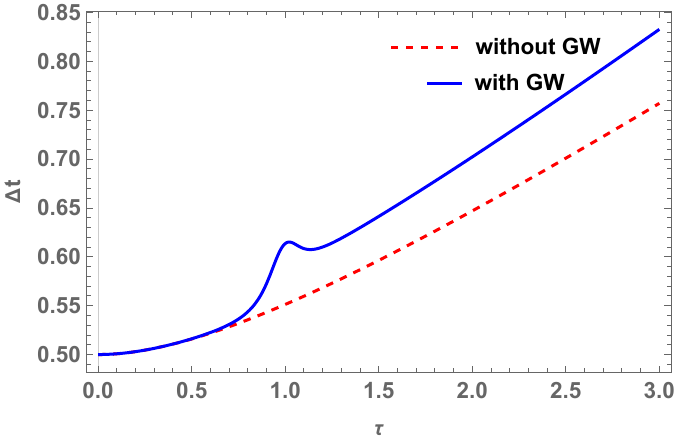}
  \caption{Deviation in $t$ co-ordinate difference}
  \label{fig:sub2}
\end{subfigure}%

\begin{subfigure}{.5\textwidth}
  \centering
  \includegraphics[width=0.9\linewidth]{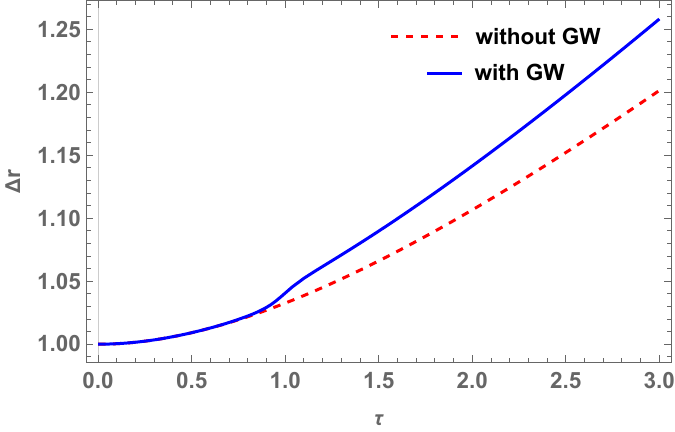}
  \caption{Deviation in $r$ co-ordinate difference}
  \label{fig:sub2}
\end{subfigure}%
\begin{subfigure}{.5\textwidth}
  \centering
  \includegraphics[width=0.9\linewidth]{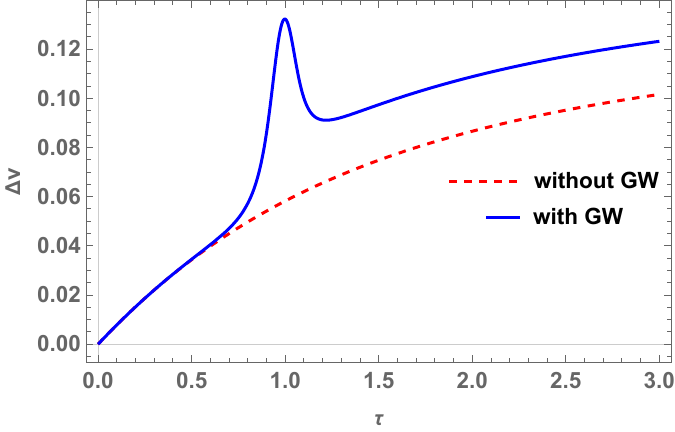}
  \caption{Deviation in velocity($v=d(\Delta r)/d\tau$) difference}
  \label{fig:sub2}
\end{subfigure}%
\caption{Memory effect in Braneworld Wormhole with $p=0.01$}
\label{fig:brw}
\end{figure}

\noindent The displacement and velocity memory effects are shown in figure \ref{fig:brw}.

\begin{figure}[hbt!]
\centering
\begin{subfigure}{.5\textwidth}
  \centering
   \includegraphics[width=0.9\linewidth]{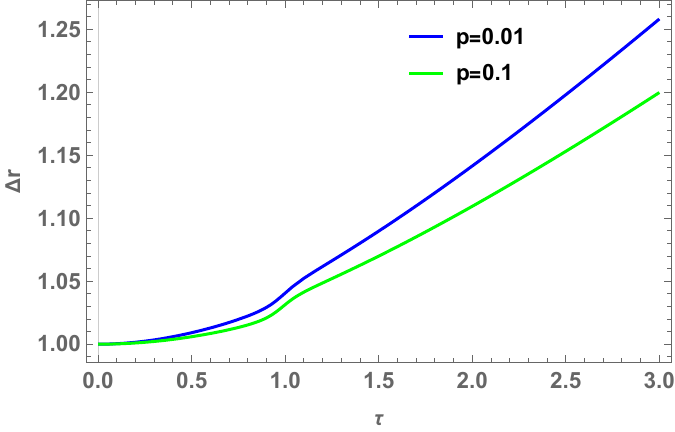}
  \caption{Comparison of $r$ co-ordinate difference}
  \label{fig:sub1}
\end{subfigure}%
\begin{subfigure}{.5\textwidth}
  \centering
  \includegraphics[width=0.9\linewidth]{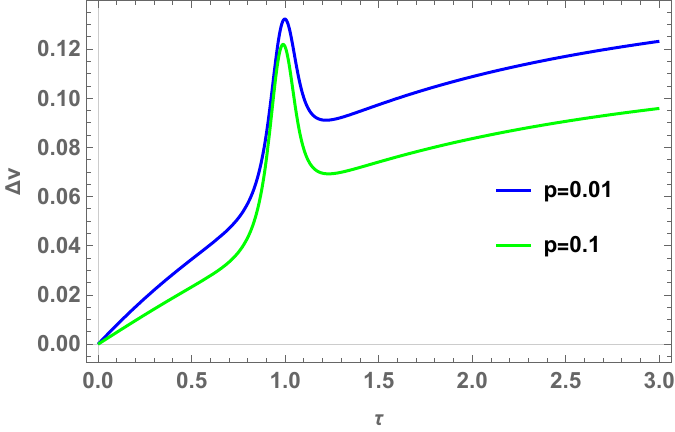}
  \caption{Comparison of velocity($v=d(\Delta r)/d\tau$) difference}
  \label{fig:sub2}
\end{subfigure}
\caption{Comparison between different values of the metric parameter $p$ in Braneworld Wormhole}
\label{fig:brwcom}
\end{figure}

Dependence of the memory effects on the wormhole hair $p$ are depicted in figure \ref{fig:brwcom} for two different values of $p$.
%\newpage
%\subsection{Black holes in Pure Lovelock Gravity}

\section{Comparison of Memory effect}\label{sec5}

\begin{figure}[hbt!]
\centering
\begin{subfigure}{.5\textwidth}
  \centering
   \includegraphics[width=1\linewidth]{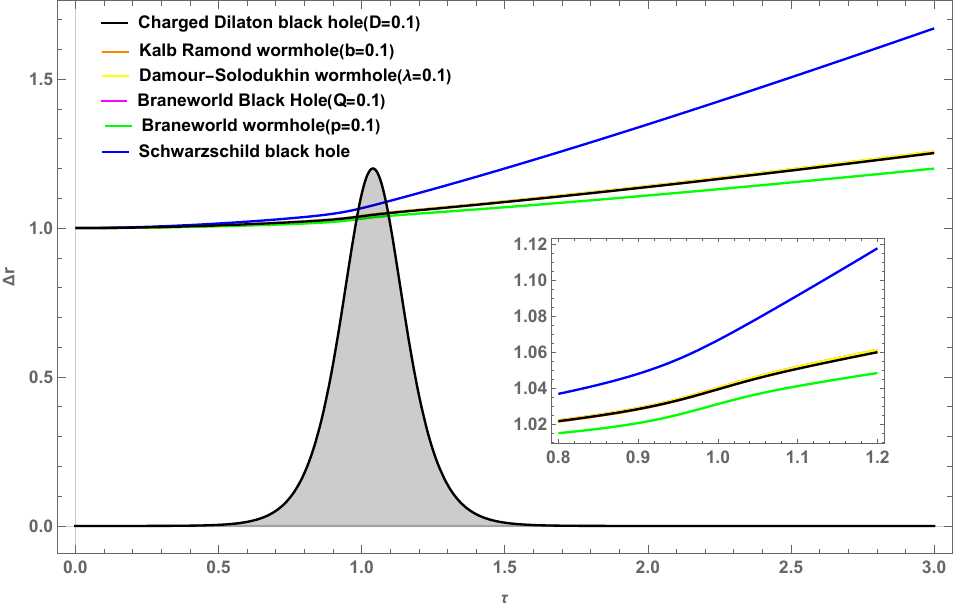}
  \caption{Comparison of $\Delta r$ }
  \label{master_r}
\end{subfigure}%
\begin{subfigure}{.5\textwidth}
  \centering
   \includegraphics[width=1\linewidth]{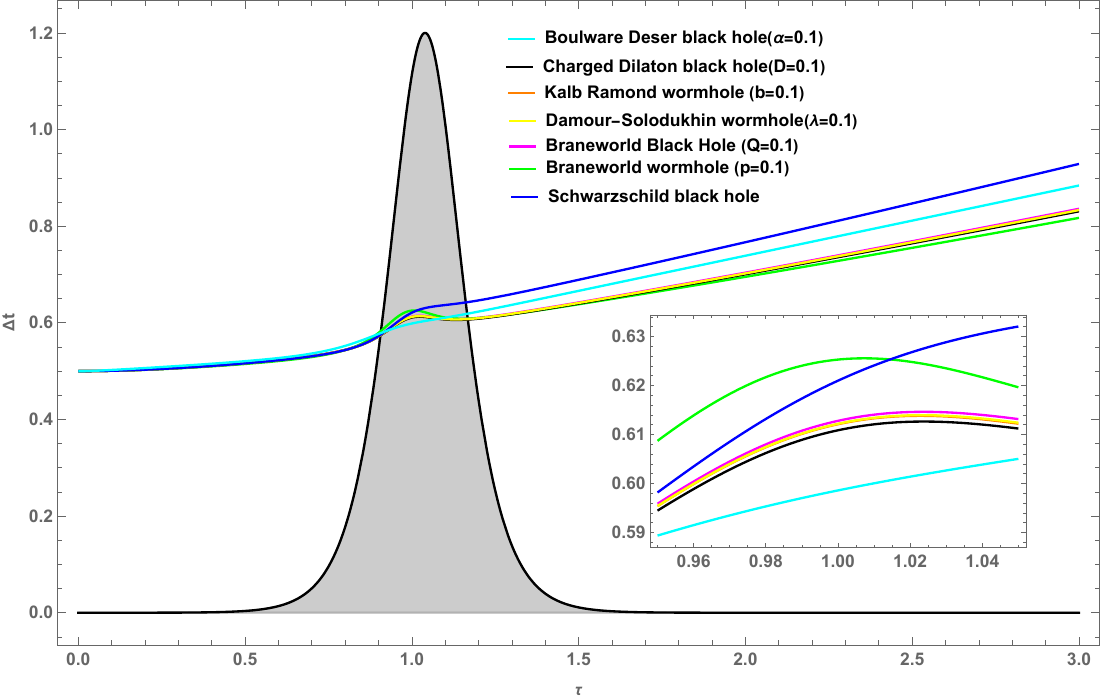}
  \caption{Comparison of $\Delta t$}
  \label{master_t}
\end{subfigure}%

\begin{subfigure}{.5\textwidth}
  \centering
  \includegraphics[width=1\linewidth]{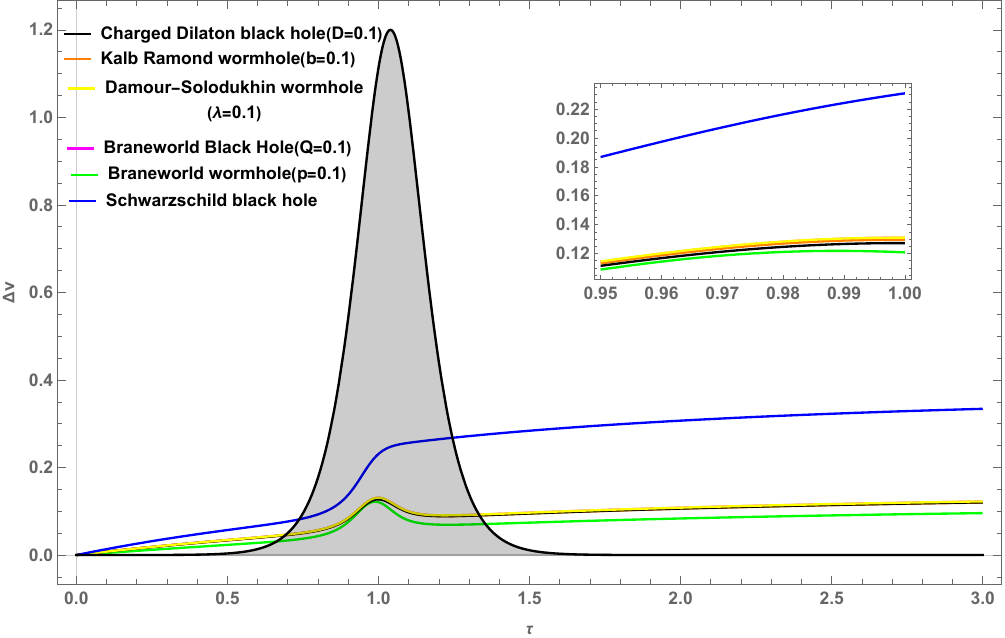}
  \caption{Comparison of $\Delta v$}
  \label{master_v}
\end{subfigure}%
\caption{Comparison of the effect of a Gravitational Wave in various spacetime geometries}
\label{fig:Master}
\end{figure}

\noindent We have studied memory effect for various static and spherically symmetric geometries, some of which represent black holes and others are wormholes. We now combine all our results in a single plot which shows that the memory effect obtained in difference geometries are quite distinct from each other. In the following plot (figure \ref{fig:Master}) we have taken specific values of the parameters but a more extensive study with a variety of parameter values can be performed in comparison to a Schwarzschild black hole to standardise the differences between those spacetime geometries.  
 In figures \ref{master_r} and \ref{master_t} we can clearly see that memory effect manifests differently for different spacetime geometries. Figure \ref{master_v} represents a comparative study of velocity memory effect for various static, spherically symmetric spacetimes.
 %and plots (a) and (c) provide an idea of how small the comparative differences really are among the geometries and how highly sensitive an instrument must be to be able to probe these differences. Nevertheless, these difference do exist and that is what we wish to point out in this study.

\section{Conclusion} \label{conclusion} \label{sec6}
\noindent With the current advancements in Gravitational Wave research and the promise of highly sensitive upcoming detectors, we now have the opportunity of studying systems and phenomena which were out of our reach before. Memory effect is one such occurrence whose detection is becoming more realistic with the improving technology of gravitational wave detectors. We hence propose Memory effect as a criterion for differentiating between various compact object geometries.\\
\indent In this paper we discuss memory effect in different static and spherically symmetric solutions of Einstein gravity as well as of theories beyond General Relativity. We, at first, have briefly discussed the geometries of various wormhole and black hole spacetimes that those spherically symmetric solutions represent. Then we have analysed the
displacement and velocity memory effects by studying
neighbouring geodesics in each of these backgrounds in
the presence of a localised GW pulse. We have shown
explicitly, how the geodesic separations evolve before and
after the passage of the pulse. This clearly establishes
the existence of both displacement and velocity memory
effect. We then compare the result with that of Schwarzschild background.\\
\indent In all our computations we have taken parameter values at least an order of magnitude lower than $1$ as we have set $M=1$. If we keep the memory effect in Schwarzschild metric as a benchmark, we can see that all the other geometries lie only on one side of the Schwarzschild plot and we expect that they should not cross the Schwarzschild benchmark curve for any positive value of the parameter (keeping in mind the standard of $M=1$ and the corresponding range of values that parameters can take under the purview of Solar system tests of GR). (Note: We have not shown Boulware-Deser in figures \ref{master_r} and \ref{master_v}  as the scale in which memory effect is manifested in this spacetime is outside the scale of all other metrics and hence it could not be included in this plot which has been set to a particular vertical scale).\\
\indent %\textcolor{red}{
One caveat regarding the use of gravitational memory effect as a measuring tool is that it can only differentiate between
different exterior geometries, for example, it cannot be used to differentiate between a black hole and a compact object having identical static and spherically symmetric geometry in the exterior. Therefore this method can only differentiate between compact objects if they give rise to different background geometry.
Most of the spacetime geometries that we have considered here are dependent on certain parameters representing black hole/wormhole hairs. We have shown through our analysis how the GW memory depends on these hairs for a wide range of values for the parameters (but small values as we fix $M=1$ everywhere). \\
%so although it can be argued that the memory effect we have obtained could be an artefact of these parameters we can deny this with high probability and conclude the physicality of memory effect because we have tried to obtain this effect for a wide range of values for the parameters (but small values as we fix $M=1$) and thus it is highly unlikely that memory effect should be a coincidence.
\indent Current ground based gravitational wave detectors, like LIGO, have a strain sensitivity of about $10^{-20}$. 
LIGO is insensitive to the memory from most sources because the detector
response timescale is generally much shorter (of the order of few milliseconds) than the rise-time for the memory signals \cite{favata}. We hope that future detectors like LISA would be the perfect setup for seeing memory effect due to the fact that it will have higher strain sensitivity (of the order of $10^{-23}$) in
the low-frequency band where typical memory sources are stronger \cite{favata, Reinhard}. LISA has a longer detector response time scale (of the order of few years) and hence has a higher chance of data accumulation \cite{Bailes, Babak}. Since it is a space-based system, it will naturally be in free-fall throughout its course. From an experimental point of view, memory effect is important because it permits a measurement to be made, not during a short burst of gravitational radiation, but over a much longer time, during which the particles can still be assumed to be free.\\
\indent Although we have considered static and spherically symmetric spacetime geometries, observational data indicates that most astrophysical systems in our universe undergo rotation. Hence a possible future goal would be to study gravitational memory effect in rotating compact objects. It would be interesting to see how the memory effect, for example, in a rotating black hole would differ from a stationary black hole as well as any other black hole or wormhole model. Also we can study the memory effect using
symmetries at null infinity using the Bondi-Sachs formalism and explore how the variation in the Bondi mass aspect, related
to the memory effect, depend explicitly on different black hole and wormhole
backgrounds used here, following the formalism used in \cite{SC4}. We hope we can address these issues in near future. 
\section{Acknowledgement}

We acknowledge Sumanta Chakraborty for initiating this project. We also thank him for various insightful comments and discussions during different stages of this project.  SG acknowledges IACS (Indian Association for the Cultivation of Science) for providing financial assistance through the Master's Fellowship. SB acknowledge DAE for providing a post-doctoral fellowship through RRF scheme (grant no: $1003/(6)/2021/{\rm RRF}/{\rm R\&D}-{\rm II}/4031, {\rm dated}: 20/03/2021$).


\begin{thebibliography}{99}
\bibitem{md}Maeda, Hideki, and Naresh Dadhich. "Matter without matter: Kaluza-Klein spacetime in Einstein-Gauss-Bonnet gravity." Physical Review D 75.4 (2007): 044007.
\bibitem{maeda} Bhattacharya, Sourav, and Sumanta Chakraborty. "Constraining some Horndeski gravity theories." Physical Review D 95.4 (2017): 044037.
\bibitem{Deserb}  D. Boulware and S. Deser, String Generated Gravity Models, Phys. Rev. Lett. 55 (1985)
2656.
\bibitem{BDpaper} Garraffo, Cecilia, and Gaston Giribet. "The Lovelock black holes." Modern Physics Letters A 23.22 (2008): 1801-1818.
\bibitem{braneworld}Kanno, Sugumi, and Jiro Soda. "Radion and holographic brane gravity." Physical Review D 66.8 (2002): 083506.
\bibitem{echoeskerr}Bueno, Pablo, et al. "Echoes of Kerr-like wormholes." Physical Review D 97.2 (2018): 024040.
\bibitem{LIGO1}  B. Abbott et al. (LIGO Scientific, Virgo), Phys. Rev.
Lett. 116, 061102 (2016).
\bibitem{LIGO2} B. P. Abbott et al. (LIGO Scientific, Virgo), Phys. Rev.
Lett. 119, 161101 (2017), arXiv:1710.05832 [gr-qc].
\bibitem{EHT1}  K. Akiyama et al. (Event Horizon Telescope), Astrophys.
J. Lett. 875, L1 (2019), arXiv:1906.11238 [astro-ph.GA].
\bibitem{EHT2} K. Akiyama et al. (Event Horizon Telescope), Astrophys.
J. Lett. 875, L2 (2019), arXiv:1906.11239 [astro-ph.IM].
\bibitem{EHT3}  K. Akiyama et al. (Event Horizon Telescope), Astrophys.
J. Lett. 875, L3 (2019), arXiv:1906.11240 [astro-ph.GA].
\bibitem{EHT4}  K. Akiyama et al. (Event Horizon Telescope), Astrophys.
J. Lett. 875, L4 (2019), arXiv:1906.11241 [astro-ph.GA].
\bibitem{EHT5} K. Akiyama et al. (Event Horizon Telescope), Astrophys.
J. Lett. 875, L5 (2019), arXiv:1906.11242 [astro-ph.GA].
\bibitem{EHT6} K. Akiyama et al. (Event Horizon Telescope), Astrophys.
J. Lett. 930, L12 (2022).
\bibitem{EHT7} K. Akiyama et al. (Event Horizon Telescope), Astrophys.
J. Lett. 930, L17 (2022).
\bibitem{Yunes}  N. Yunes and X. Siemens, Living Rev. Rel. 16, 9 (2013),
arXiv:1304.3473 [gr-qc].
\bibitem{LIGO-VIRGO} The LIGO Scientific Collaboration, The Virgo Collabo-
ration, The KAGRA Collaboration, and R. e. a. Abbott,
arxiv:2112.06861 (2021)DE13575.
\bibitem{Ohme}  N. V. Krishnendu and F. Ohme, Universe 7, 497 (2021),
arXiv:2201.05418 [gr-qc].
\bibitem{Genzel}  T. Johannsen, A. E. Broderick, P. M. Plewa, S. Chat-
zopoulos, S. S. Doeleman, F. Eisenhauer, V. L. Fish,
R. Genzel, O. Gerhard, and M. D. Johnson, Phys.
Rev. Lett. 116, 031101 (2016), arXiv:1512.02640 [astro-
ph.GA].
\bibitem{Yunes1}  D. Ayzenberg and N. Yunes, Class. Quant. Grav. 35,
235002 (2018), arXiv:1807.08422 [gr-qc].
\bibitem{psaltis} D. Psaltis, Gen. Rel. Grav. 51, 137 (2019),
arXiv:1806.09740 [astro-ph.HE].
\bibitem{SC1}  I. Banerjee, S. Chakraborty, and S. SenGupta, Phys. Rev.
D 101, 041301 (2020), arXiv:1909.09385 [gr-qc].
\bibitem{SC2}  S. Chakraborty, E. Maggio, A. Mazumdar, and P. Pani,
(2022), arXiv:2202.09111 [gr-qc].
\bibitem{SC3} A. K. Mishra, S. Chakraborty, and S. Sarkar, Phys. Rev.
D 99, 104080 (2019), arXiv:1903.06376 [gr-qc].
\bibitem{SC4} I. Chakraborty, S. Bhattacharya and S. Chakraborty, Phys. Rev. D 106, 104057, (2022), arXiv: 2207.00226.
\bibitem{Nichols1}  O. M. Boersma, D. A. Nichols, and P. Schmidt, Phys. Rev. D 101, 083026 (2020).
\bibitem{braginsky} V. B. Braginsky and L. P. Grishchuk, Sov. Phys. JETP 62, 427 (1985).
\bibitem{favata} M. Favata, Class. Qtm. Grav. 27, 084036 (2010).
\bibitem{braneBH}Dey, Ramit, Sumanta Chakraborty, and Niayesh Afshordi. "Echoes from braneworld black holes." Physical Review D 101.10 (2020): 104014.
\bibitem{RS1}Randall, Lisa, and Raman Sundrum. "Large mass hierarchy from a small extra dimension." Physical review letters 83.17 (1999): 3370.
\bibitem{RS2}Randall, Lisa, and Raman Sundrum. "An alternative to compactification." Physical Review Letters 83.23 (1999): 4690.
\bibitem{RS3}Maartens, Roy, and Kazuya Koyama. "Brane-world gravity." Living Reviews in Relativity 13 (2010): 1-124.
\bibitem{gauss1}Shiromizu, Tetsuya, Kei-ichi Maeda, and Misao Sasaki. "The Einstein equations on the 3-brane world." Physical Review D 62.2 (2000): 024012.
\bibitem{gauss2}Harko, T., and M. K. Mak. "Vacuum solutions of the gravitational field equations in the brane world model." Physical Review D 69.6 (2004): 064020.
\bibitem{gauss3}Aliev, A. \& Gumrukcuoglu, A.. (2005). Charged Rotating Black Holes on a 3-Brane. Physical Review D. 71. 10.1103/PhysRevD.71.104027. 
\bibitem{gauss4}Chakraborty, Sumanta, and Soumitra SenGupta. "Spherically symmetric brane in a bulk of f (R) and Gauss–Bonnet gravity." Classical and Quantum Gravity 33.22 (2016): 225001.
\bibitem{gauss5}Chakraborty, Sumanta, and Soumitra SenGupta. "Spherically symmetric brane spacetime with bulk f (R) gravity." The European Physical Journal C 75.1 (2015): 11.
\bibitem{CD}David Garfinkle, Gary T. Horowitz, and Andrew Strominger
Phys. Rev. D 43, 3140 – Published 15 May 1991; Erratum Phys. Rev. D 45, 3888 (1992)
\bibitem{brag} Braginskij, V.B., \& Grishchuk, L.P. (1985). Kinematic resonance and the memory effect in free mass gravitational antennas. Zhurnal Ehksperimental'noj i Teoreticheskoj Fiziki, 89(3), 744-750.
\bibitem{damsol}Damour, Thibault, and Sergey N. Solodukhin. "Wormholes as black hole foils." Physical Review D 76.2 (2007): 024016.
\bibitem{lorentzianwh}M. Visser, “Lorentzian wormholes: From Einstein to Hawking,” Woodbury, USA:
AIP (1995) 412 p
\bibitem{ssgpaper}Kar, Sayan, Soumitra SenGupta, and Saurabh Sur. "Static, spherically symmetric solutions, gravitational lensing, and perihelion precession in Einstein-Kalb-Ramond theory." Physical Review D 67.4 (2003): 044005.
\bibitem{kr}Chakraborty, Sumanta, and Soumitra SenGupta. "Strong gravitational lensing—A probe for extra dimensions and Kalb-Ramond field." Journal of Cosmology and Astroparticle Physics 2017.07 (2017): 045.
\bibitem{Bailes} M. Bailes et al. Gravitational-Wave physics and
astronomy in the 2020s and 2030s,  Nature Reviews Physics volume 3, pages344–366 (2021). 
\bibitem{Reinhard} R. Reinhard, LISA – Detecting and Observing
Gravitational Waves, esa bulletin 103 — august 2000. 
\bibitem{Babak} S. Babak, M. Hewitson, A. Petiteau, LISA Sensitivity and SNR Calculations, arXiv:2108.01167 [astro-ph.IM]. 


\end{thebibliography}
\end{document}